\documentclass{aastex}
\usepackage{emulateapj5}
\begin{document}
\newcommand{\Halpha}{H$\alpha$ }
\newcommand{\etal}{\mbox{et al.}}
\newcommand{\NHoo}{N_{\rm H}}
\newcommand{\NHxo}{N_{\rm x}}
\newcommand{\NHIo}{N_{\rm HI}}
\newcommand{\NHtwoone}{N$_{\rm 21cm}$ }
\newcommand{\NHII}{N_{\rm HII}}
\newcommand{\NHgx}{N_{\rm G}}
\newcommand{\acm}{ cm$^{-2}$ }
\newcommand{\as}{s$^{-1}$}
\newcommand{\Htwo}{H$_{2}$}
\newcommand{\LHB}{{\sc lhb}}
\newcommand{\ISM}{{\sc ism}}
\newcommand{\LISM}{{\sc lism}}
\newcommand{\MER}{{\sc mer}}
\newcommand{\ASCA}{{\it ASCA}}
\newcommand{\IRAS}{{\it IRAS}}
\newcommand{\ROSAT}{{\it ROSAT}}
\newcommand{\COBE}{{\it COBE}}
\newcommand{\DIRBE}{{\it DIRBE}}
\newcommand{\EUVE}{{\it EUVE}}
\newcommand{\pp}{\phn}
\newcommand{\ppp}{\phn\phn}
\newcommand{\pppp}{\phn\phn\phn}
\newcommand{\pq}{\,$\pm$\,}
\newcommand{\pd}{\phn\phn\phn\,---}
\newcommand{\plong}{\hspace{10pt}}
\newcommand{\gte}{$\infty$\phn}
\newcommand{\ZY}{0.3,1.0}
\newcommand{\ZZ}{0.5,1.0}
\newcommand{\Msun}{$M_{\odot}$}
\newcommand{\ayr}{y$^{-1}$}
\newcommand{\ARAA}[2]{ARA\&A, #1, #2}
\newcommand{\ApJ}[2]{ApJ, #1, #2}
\newcommand{\ApJL}[2]{ApJL, #1, #2}
\newcommand{\ApJSS}[2]{ApJS, #1, #2}
\newcommand{\ApJinp}{ApJ, in press}
\newcommand{\ApJsub}{ApJ, submitted}
\newcommand{\AandA}[2]{A\&A, #1, #2}
\newcommand{\AandASS}[2]{A\&AS, #1, #2}
\newcommand{\AJ}[2]{AJ, #1, #2}
\newcommand{\BAAS}[2]{BAAS, #1, #2}
\newcommand{\ASP}[2]{ASP Conf.\ Ser., #1, #2}
\newcommand{\JCP}[2]{J.\ Comp.\ Phys., #1, #2}
\newcommand{\MNRAS}[2]{MNRAS, #1, #2}
\newcommand{\N}[2]{Nature, #1, #2}
\newcommand{\PASJ}[2]{PASJ, #1, #2}
\newcommand{\PASP}[2]{PASP, #1, #2}
\newcommand{\PRD}[2]{Phys.\ Rev.\ D, #1, #2}
\newcommand{\PRL}[2]{Phys.\ Rev.\ Lett., #1, #2}
\newcommand{\RPP}[2]{Rep.\ Prog.\ Phys., #1, #2}
\newcommand{\ZA}[2]{Z.\ Astrophs., #1, #2}
\newcommand{\tenup}[1]{\times 10^{#1}}
\newcommand{\pz}[0]{\phantom{0}}

\title{Extracting the Dark Matter Profile of a Relaxed Galaxy Cluster}
\author{J.S.\ Arabadjis\altaffilmark{1}, M.W.\ Bautz\altaffilmark{1}, and
G.\ Arabadjis\altaffilmark{2}}
\altaffiltext{1}{Center for Space Research, Massachusetts Institute of
Technology, Cambridge, MA 02139; {\tt jsa@space.mit.edu},
{\tt mwb@space.mit.edu}}
\altaffiltext{2}{Mitre Corporation, 202 Burlington Road, Bedford, MA 01730;
{\tt gus@mitre.org}}

\begin{abstract} 

Knowledge of the structure of galaxy clusters is essential for an understanding
of large scale structure in the universe, and may provide important clues to
the nature of dark matter.  Moreover, the shape of the dark matter distribution
in the cluster core may offer insight into the structure formation process.
Unfortunately, cluster cores also tend to be the site of complicated
astrophysics.  X-ray imaging spectroscopy of relaxed clusters, a standard
technique for mapping their dark matter distributions, is often complicated by
the presence of cool components in cluster cores, and the dark matter profile
one derives for a cluster is sensitive to assumptions made about the
distribution of this component.  In addition, fluctuations in the temperature
measurements resulting from normal statistical variance can produce results
which are unphysical.  We present here a procedure for extracting the dark
matter profile of a spherically symmetric, relaxed galaxy cluster which deals
with both of these complications.  We apply this technique to a sample of
galaxy clusters observed with the {\it Chandra X-ray Observatory}, and comment
on the resulting mass profiles.  For some of the clusters we compare their
masses with those derived from weak and strong gravitational measurements.

\end{abstract}


\section{Introduction} 

The cold dark matter (CDM) paradigm of modern cosmology has enjoyed spectacular
success in describing the formation of large-scale structure in the universe
\citep{nfw_seven,moore_b,lahav,peacock}.  There are, however, several nagging
inconsistencies between the results of numerical CDM experiments and
observations.  On small scales, the dark matter halos in dwarf and low
surface brightness galaxies are much less cuspy than in CDM simulations
\citep{burkert,mcgaugh,moore_b}.  Disk galaxies produced in simulations tend to
have inadequate masses and angular momenta \citep{navarro}.  The number of
Milky Way satellites appears to be at least an order of magnitude lower than
CDM predictions \citep{kauffman,moore_a,klypin}.  On larger scales, some
studies \citep{tyson,smail} report galaxy clusters with central density
profiles that are flatter than CDM predictions, although these are somewhat
controversial \citep{broadhurst,shapiro}.

The density profile of bound structures which form through the hierarchical
assembly of smaller structures is usually parameterized as a power law at small
scales and a separate power law on large scales (e.g.\ \citet{jing}):
\begin{equation}
\rho(r) = \frac{\rho_0}{(r/r_s)^\alpha (1+r/r_s)^{\gamma-\alpha}}
\label{eq01}
\end{equation}

\noindent The four parameters in this description are the density $\rho_0$ at
some fiducial radius, the inner power law index $\alpha$, the outer
power law index $\gamma$, and the scale radius $r_s$ setting the break between
the two power laws.  While it is generally agreed that $\gamma=3$, the value
of $\alpha$ has generated considerable debate.  Simulations predict a value
between 1.0 \citep{nfw_six,nfw_seven} and 1.5 \citep{moore_b,fukushige},
roughly independent of halo mass and formation epoch.  In nature, however,
$\alpha$ shows a larger variation, and is likely a function of halo mass.
H$\alpha$ rotation curves of low-surface brightness galaxies indicate density
profiles which are significantly flatter -- $\alpha\sim 0.5$ -- than CDM
predictions \citep{swaters,dalcanton,borriello,swaters_new}.  X-ray
observations of galaxy clusters generally show steeper profiles, however, with
$\alpha\sim 1.2$ \citep{lewis_a} to 1.9 \citep{abg}.

These discrepancies are often ascribed to limitations of the astrophysics or
the physics included in the simulations.  Baryon physics, if included, may be
tacked on at the conclusion of a simulation according to a set of semi-analytic
and/or empirical prescriptions.  It is likely that baryon physics will play a
significant role in the evolution of the central halo.  Reports of a halo
``entropy floor'' \citep{ponman,lloyddavies} suggest non-gravitational sources
of heating and feedback either prior to or during halo formation
\citep{balogh,loewenstein,wuetal} which are probably baryonic in origin (see
\citet{mushotzky}, however).  The question then becomes one of determining
where baryon physics {\it ceases} to be important.  While the inclusion of
baryon astrophysics in sufficient detail may remedy these problems, its effects
will require a great deal of effort to disentangle \citep{frenk}.

It could be, however, that the missing ingredients in the simulations are not
all astrophysical.  One possibility is that the initial power spectrum of the
primordial fluctuations is not scale invariant.  If the primordial spectral
index of density perturbations is not precisely 1 (as is normally assumed by
appealing to standard inflationary cosmology), the formation epoch of halos may
be delayed sufficiently to ameliorate the central density problem
\citep{alam,zentner}.
Another possibility is that important dark matter particle physics is being
overlooked, and that the assumption of no non-gravitational self-interactions
is faulty.  Proposed modifications of CDM include, though are not limited to,
self-interacting dark matter \citep{spergel,firmani}, warm dark matter
\citep{hogan}, annihilating dark matter \citep{kaplinghat}, scalar field dark
matter \citep{hu,goodman}, and mirror matter \citep{mohapatra}, each of which
is invoked to soften the core density profile.  Many of these modifications
will soften the core profile of galaxy clusters as well, in conflict with many
X-ray determinations of mass profiles, although other astrophysical processes
such as the adiabatic contraction of core baryons \citep{hennawi} may mitigate
this effect.

In an effort to discriminate between CDM modifications and other astrophysical
influences we are mapping the dark matter profiles of a large sample of galaxy
clusters.  Specifically, we use imaging spectroscopy from the
{\it Chandra X-ray Observatory} \citep{weisskopf} to determine the deprojected
temperature and density profiles of the baryonic content of each galaxy
cluster, which we then use to derive its dark matter profile.  In this paper we
describe our method, and apply it to a sample of low- and moderate-redshift
clusters.  We describe our spectral deprojection technique in
\S\ref{depro}; we discuss the problems involved in converting these results to
a mass profile and our solution in \S\ref{mass}; we examine the effects that
cooling flow model assumptions have on our profiles, and we present a
statistical analysis of the models and a prescription for choosing among them
using Markov Chain Monte Carlo sampling in \S\ref{mcmc}.  Finally, we summarize
our findings in \S\ref{summary}.  In a subsequent paper we will examine these
profiles for their implications for large scale structure formation and dark
matter particle properties (Arabadjis \& Bautz, in preparation [AB]).

\section{Spectral deprojection} \label{depro} 

We begin with a {\it Chandra} imaging spectroscopic observation of a galaxy
cluster using the ACIS detector, either with the S3 chip or the I array.  We
start with a level 2 data set that has been processed in the usual way,
filtered for periods of high background using the procedures described in
the CIAO Science
Threads\footnote{See {\tt http://asc.harvard.edu/ciao/threads/all.html}},
with point sources removed (for details see \citet{abg}).
After locating the center of the projected emissivity, we lay down a series of
adjacent, concentric annuli centered on the emission peak.  The annular
dimensions are set to include enough source photons (1000-2000+) to reliably
determine the plasma temperature.  RMF and ARF response matrices are
constructed, as is a background spectrum from a region external to the
outermost annulus.  The spectra are recorded in PI format, and grouped such
that there is a minimum of 20 counts per channel.

Our spectral deprojection has been presented elsewhere \citep{abg}, and so we
just briefly summarize here.  To derive spherical radial profiles we construct
a model consisting of $N$ concentric spherical shells whose inner and outer
radii correspond to the inner and outer cylindrical radii of the projected
annuli in the data set.  The volume intersection matrix ${\sf V}$, whose
elements $V_{ij}$ contain the volume of spherical shell $j$ intersected by a
cylindrical shell formed by the projection of annulus $i$, is used to set the
linear relations between each of the normalizations as specified by the binning
geometry.  Each shell $i$ on $[1,N]$ (1 is the innermost shell -- a sphere --
and $N$ is the outermost shell) contains an optically thin thermal plasma whose
emission characteristics are determined by the MEKAL model
\citep{meweg,mewel,kaastra_mekal,liedahl} within XSPEC \citep{arnaud} using two
free parameters, the temperature $T$ and the normalization $K$.

In some models we will allow the innermost $N_c$ shells to contain a second
emission component at a (lower) temperature $T_c$ as a first-order treatment
of the cooler plasma.  This cool component is assumed to be in pressure
equilibrium with the hot component; this means that the cool and hot components
cannot both be in hydrostatic equilibrium.  We {\it assume} that the hot
component is in hydrostatic equilibrium.  We adopt this form for the cooling
flow model for two reasons: (1) there is no evidence that the plasma in cooling
flow cluster cores cools below about 1 keV \citep{peterson} as in the isobaric
cooling flow model of \citet{mushotzky_szym}, and (2) it is arguably the
simplest adjustment that can be made to the uniphase model.  The cool component
could be arranged in droplets or filaments which are replenished as they
migrate to the unspecified sink at $r=0$, effecting a hydrodynamical
equilibrium.  The details of the geometry of the cool component are unimportant
since they are below our spatial resolution limit; it is only our assumption
that it is in pressure equilbrium with the hot component which has
observational consequences.  In this study we use the pressure gradient in the
core to measure the radial dependence of the enclosed gravitating mass.

The number of parameters in this model is rather large.  Each MEKAL component
contains six parameters, of which two ($T$ and $K$) are allowed to vary.
Thus the $N$ annuli in the data set are modelled using $N+N_c$ emission
components.  Including a Galactic absorption column yields a model with
$6(N+N_c)+1$ parameters (although only a subset of these actually vary).  Each
of the $N$ annuli independently constrain between 1 and $N$ of the model shells
in the fitting, and so the complete XSPEC model contains $N[6(N+N_c)+1]$
components, $2(N+N_c)+1$ of them variable.  The current version of XSPEC admits
models with 1000 parameters, any 100 of which can vary.  This limits our model
to $N=12$ annuli for $N_c=0$ (876 parameters, 25 variable) and $N=10$ annuli
for $N_c=2$ (730 parameters, 25 variable).  We have written a program which
reads in the data annuli dimensions and writes out an XSPEC script that handles
all of the data manipulation.  The script initializes the model parameters,
performs the $\chi^2$ minimization, and calculates parameter uncertainties.
This software is available to the public through requests to the authors.

Most spectral deprojection schemes rely on an ``onion peeling'' approach
\citep{fabian_etal,allen_fabian,david,lewis_b,sun}: the outermost annulus is
modeled using the outermost spherical shell; its model parameters are then
frozen and its emission is subtracted from all annuli interior to it.  The next
most outer shell is then modelled, the resulting model again frozen and
subtracted from the interior, and so forth, until the entire cluster has been
modelled.  The virtue of this technique is that the number of model components
scales as $N$ instead of $N^2$, allowing for greater spatial detail.  However,
because the parameters of each model shell are frozen and the model is
subtracted from interior shells as if it contained zero uncertainty, the
technique does not find the global ``best fit'' of the model parameters.  In
many cases, the errors quoted in onion-peeling analyses are the uncertainties
associated with a single layer of the onion \citep{david}.  Error estimates
derived from Monte Carlo simulations are more reliable, limited primarily by
the number of simulations used to estimate the uncertainties \citep{lewis_b}.
In our study all of the model parameters are fit simultaneously; the subsequent
determination of error in subsets of interesting variables is a true expression
of the parameter uncertainties in the model.  We note here that although our
model appears to have many more parameters than the other techniques, in
actuality the numbers are equivalent because of the web of linear dependences
among the component normalizations.

\section{Mass profiles} \label{mass} 

Many deprojection methods rely on analytic formulae for the radial run of
temperature, surface brightness or mass, either during or after the fitting
process \citep{allen_etal,david,hicks,pizzolato}.  The greatest advantage of
a parametric treatment is numerical stability.  In addition, it is common
practice to smooth noisy profiles before using them in subsequent calculations.
This latter technique is especially useful when deriving gravitating mass
profiles since a numerical derivative must be computed.  A serious drawback of
these approaches is that it is difficult to quantify the effect of the
parameterization, or the smoothing, on the results.  Additionally, it is often
difficult or impossible to propagate errors through to the results.

Our non-parametric deprojection technique does not guarantee smooth temperature
and density profiles, so we have devised a method whereby the mass profile is
computed from within the error envelope of $(\rho(r),T(r))$.  By choosing a
statistically reasonable realization of the model, we are able to compute mass
profiles which are not only smoother than those obtained from the unconstrained
$(\rho,T)$ set, but which avoid the unphysical results that arise due to the
statistical fluctuations inherent in measurements.  Essentially, our procedure
imposes physically motivated constraints to reduce the uncertainty in the
temperature and mass profiles, and provides a statistic which characterizes the
reliability of the mass reconstruction.

The standard procedure for extracting the gravititating mass profile of a
galaxy cluster is to insert its deprojected temperature and density profiles
into the hydrostatic equation \citep{sarazin}:
\begin{equation}
M_r = -\frac{kT}{G\mu m_p/r}
\left( \frac{d\log{T}}{d\log{r}} +\frac{d\log{\rho}}{d\log{r}} \right) \, ,
\label{eq02}
\end{equation}

\noindent Here $T$ and $\rho$ are the local (baryonic) plasma temperature and
density, $r$ is the spherical radius, $M_r$ is the total mass enclosed within
$r$ (i.e.\ baryons plus dark matter), and $m_p$ and $\mu$ are the proton mass
and mean particle weight, respectively.  Implicit here is the assumption that
the cluster is supported solely by an isotropic thermal pressure gradient, i.e.
that random motions greatly exceed the bulk rotational motion, and that the
magnetic field energy density is negligible in comparison with the thermal
energy content of the plasma.  We further assume that no recent merger event
has caused a disruption in the pressure and density.  Given the run of density
and temperature in our binning scheme
($r_i$, $\rho_i$, and $T_i$, $i=1,2,...,N$), we can calculate $M_i$ using a
simple difference equation version of Equation~\ref{eq02}:
\begin{equation}
M_i = -A \, r_i \, T_i
\left(\frac{t_{i+1}-t_{i-1}+d_{i+1}-d_{i-1}}{x_{i+1}-x_{i-1}}\right)
\label{eq03}
\end{equation}

\noindent where $A = k/G \mu m_p$, $x_i = \log_{10}{(r_i)}$,
$d_i = \log_{10}{(\rho_i)}$, and $t_i = \log_{10}{(T_i)}$.  Since the goal of
this technique is to derive enclosed mass profiles, we define $r$ as the outer
radius of each shell.

Because of measurement error a mass profile calculated in this way is not
guaranteed to be physically reasonable.  Even if the assumptions of spherical
symmetry and hydrostatic equilibrium were valid, statistical fluctuations in
the temperature measurements could result in unphysical points in our derived
mass profile, for example $dM_r/dr<0$  or even $M_r<0$.  To deal with the
non-physical fluctuations, we proceed under the assumptions that all unphysical
values in the mass profile are due to measurement uncertainty in either $\rho$
or $T$.  Specifically, to estimate the $(\rho,T)$ profiles we impose the
condition that the computed total gravitating mass profile is consistent with
\begin{equation}
\rho({\rm total}) \, = \, \rho({\rm baryons}) + \rho({\rm dark \,\, matter}) \,
\ge 0 \hspace{10pt} {\rm for} \hspace{10pt} r \ge 0
\label{eq04}
\end{equation}

\noindent Since $\rho(total) = \frac{1}{4\pi r^2} \frac{d M_r}{d r}$ and
$1/4\pi r^2$ is positive definite, this is equivalent to the constraint
\begin{equation}
dM_r/dr \ge 0
\label{eq05}
\end{equation}

\noindent Combining Equation~\ref{eq05} with the boundary condition
$M_r(0)\ge0$ (e.g., allowing for the presence of an unresolved central object)
we obtain a second constraint:
\begin{equation}
M_r(r) \ge 0
\label{eq06}
\end{equation}

Perhaps the most natural way to impose these conditions would be to invoke them
as a Bayesian prior in the spectral fitting procedure.  Bayes' Theorem
\citep{bayes,papoulis} states that the probability of model ${\sf M}$ given
data set ${\sf D}$ (the posterior distribution $P({\sf M|D})$) is proportional
to the product of the probability of that data set given the model (the
likelihood function $P({D|M})$) and the probability of the model itself (the
prior knowledge function $P({\sf M})$).
\begin{equation}
P({\sf M}|{\sf D}) \propto 
P({\sf D}|{\sf M}) \cdot
P({\sf M})
\label{eq07}
\end{equation}

\noindent The model prior could be chosen to enforce the constraints in
Equations~\ref{eq05} and \ref{eq06}.  Thus for certain combinations of model
parameters $T_i, \rho_i$, we could choose a prior such that
$P({\sf M}(T_i,\rho_i))=0.$

Unfortunately this approach is computationally impractical.  The difficulty
lies in the fact that the relatively simple constraints on $M(r)$ and $dM/dr$
lead to complicated (nonlinear) constraints of the coupled parameters of the
spectral models $(K_i,T_i)$.  The problem is more tractable if we instead apply
the Bayesian constraints after unconstrained density and temperature profiles
have been determined using the spectral deprojection method described in
\S\ref{depro}.  Thus we compute an unconstrained mass profile from
Equation~\ref{eq03}, and check it for consistency with the constraints in
Equations~\ref{eq05} and \ref{eq06}.  These three equations yield constraints
on the solutions to the difference equations which we define using $Y_i$ and
$Z_i$:
\begin{equation}
Y_i = -(t_{i+1} - t_{i-1} + d_{i+1} - d_{i-1}) \ge 0
\label{eq08}
\end{equation}

\vspace{-11pt}
\begin{eqnarray}
Z_i & = &
-r_{i+1} T_{i+1}
\left(\frac{t_{i+2} - t_{i  } + d_{i+2} - d_{i  }}{x_{i+2}-x_{i  }} \right)
\nonumber \\
 & + & r_{i} T_{i}
\left(\frac{t_{i+1} - t_{i-1} + d_{i+1} - d_{i-1}}{x_{i+1}-x_{i-1}} \right)
 \ge 0
\label{eq09}
\end{eqnarray}

\noindent Physically these constraints prohibit the pressure of the X-ray
emitting plasma from rising with radius.  Using the model parameter error
estimates as a guide, we allow the temperature and normalization of each model
component to vary if either of these constraints is violated by the profile.
Our goal is thus to find a new set of density and temperature values
$\tilde{\rho}$ and $\tilde{T}$ which are as close as possible to the the
original profiles (``fidelity'') but which obey the constraints of
Equations~\ref{eq08} and \ref{eq09} (``physicality'').

At first glance this appears to be a standard problem in constrained
optimization.  Two features of the problem suggest than an alternate route is
preferable, however.  First, as mentioned above, the constraints are non-linear
functions of the independent variables, so the constrained optimization
approach is very complex.  Second, if we exercise control over the competing
interests of fidelity and physicality, as incorporated in a penalty function,
rather than simply eliminating solutions which violate Equations~\ref{eq08} and
\ref{eq09}, we retain the ability to modulate departures from the unconstrained
profiles.  For example, it may be the case that the temperature profile would
need substantial alteration in order to satisfy the rigid imposition of
constraints~\ref{eq08} and \ref{eq09}, but that a minor violation of
Equation~\ref{eq09} (say, at only one point in the mass profile) would allow a
temperature profile of much greater fidelity.  In this instance it is to our
advantage to have the ability to set the relative weighting between the
fidelity and physicality terms in the penalty function.

Implicit in our choice of a penalty function to represent the constraints is
the assumption that our estimate of the location of the peak of the model
probability distribution function $P_{\rm est}({\sf M}(T_i,\rho_i))$, obtained
by enforcing fidelity, is close to the actual peak of the distribution
$P_{\rm 0}({\sf M}(T_i,\rho_i))$, which would obtain if the Bayesian priors
were utilized during the spectral deprojection.  This idea is shown
schematically in Figure~\ref{f01}.  As the figure illustrates, in the absence
of significant small-scale structure in $P({\sf M}(T_i,\rho_i))$,
$P_{\rm est}({\sf M}(T_i,\rho_i))$ will ideed lie near
$P_{\rm 0}({\sf M}(T_i,\rho_i))$ in parameter space.  We therefore cast our
physicality constraints as terms in a cost function, and henceforth refer to
the optimized solution for $M$ as the {\it constrained profile}, with the
understanding that it is an estimate of the fully Bayesian solution, and that
the constraints do not rigorously forbid excursion into disfavored regions of
parameter space.  Our formulation of the cost function will reflect our
subjective assessment of the relative importance of fidelity versus
physicality, and use standard optimization algorithms to minimize it.

We use the likelihood to characterize the fidelity term in the cost function.
The probability density $P$ at the vector $(\tilde{\rho},\tilde{T}) =
(\tilde{\rho_1},\tilde{\rho_2},...,\tilde{\rho_N},\tilde{T_1},\tilde{T_2},...,
\tilde{T_N}$), which is near the unconstrained profiles $(\rho,T)$, is given by
\begin{equation}
P = (2\pi)^{N/2} \, \prod^{N}_{i=1} \, (\sigma_{\rho_i} \sigma_{T_i})^{-1} \,
\, e^{-(\tilde{\rho_i}-\rho_i)^2/2 \sigma_{\rho_i}^2} \,
\, e^{-(\tilde{T_i}-T_i)^2/2 \sigma_{T_i}^2}
\label{eq10}
\end{equation}

\noindent where we have assumed that the errors in $\rho$ and $T$ are
uncorrelated and normally distributed, characterized by $\sigma_{\rho_i}$ and
$\sigma_{T_i}$, respectively.\footnote{In principle it would be more accurate
to make use of the fact that the luminosity of a shell $i$,
$L_i \propto \rho_i^2 \, T_i^{1/2}$, is tightly constrained by the
observations, and therefore that deviations in $d$ and $t$ are correlated:
$\delta d = -\frac{\delta t}{4}$.  In practice, however, this is unnecessary
since the relative uncertainty in the temperature of a shell is enormous
compared with the uncertainty in its density.}  In maximum likelihood methods,
cost functions are conveniently defined as the negative log of the likelihood
(e.g.\ \citet{vonmises}), so we define the fidelity penalty function $Q$ as
\begin{eqnarray}
Q & = & -2\log P \nonumber \\
  & = & N \log(2\pi) + \nonumber \\
  &   & 2 \sum^{N}_{i=1} \, \left[ \, \log\sigma_{\rho_i} + \log\sigma_{T_i} +
\frac{(\tilde{\rho_i}-\rho_i)^2}{2\sigma_{\rho_i}^2} +
\frac{(\tilde{T_i}-T_i)^2}{2\sigma_{T_i}^2} \, \right] \nonumber \\
  & = & Q_0 + \chi^2
\label{eq11}
\end{eqnarray}

\noindent Here $Q_0$ is a constant that depends only on the measurement errors
and $\chi^2$ is the $2N$-dimensional variance:
\begin{eqnarray}
\chi^2 & = &
\sum^{N}_{i=1} \, \left[ \, 
\frac{(\tilde{\rho_i}-\rho_i)^2}{\sigma_{\rho_i}^2} +
\frac{(\tilde{T_i}-T_i)^2}{\sigma_{T_i}^2} \, \right] \nonumber \\
       & = & \chi^2_{\rho} + \chi^2_{T}
\label{eq12}
\end{eqnarray}

\noindent We will ignore $Q_0$ since it will not affect the minimization.

We follow a similar procedure to derive the physicality terms in the
cost function.  We wish to incorporate the physicality contraints by penalizing
negative values of $Y_i$ and $Z_i$ in Equations~\ref{eq08} and \ref{eq09}
without attempting to maximize them if they are positive.  Consider the penalty
function
\begin{eqnarray}
g(x) & = & \left(|x/x_0|-x/x_0\right)^\eta +
         \left(|x/x_0-1|+x/x_0-1\right)^\eta \nonumber \\
     & - & \ln\left(\onehalf\right) + \ln(C) \hspace{30pt} (\eta > 1)
\label{eq13}
\end{eqnarray}

\noindent This function has several useful properties:  (1) it has a value of
zero for $0\le x\le x_0$; (2) it increases rapidly for $x<0$ or $x>x_0$; (3) it
has a continuous first derivative everywhere for $\eta\ge 2$ (or everywhere
but $x=0$ or $x=x_0$ for $1<\eta<2$), making it well-suited for optimization
routines that rely on gradient information to increase their efficiency; and
(4) $h(x) = e^{-g(x)}$ is a well-behaved probability distribution function.
$C$ is of order one; for arbitrary values of $\eta>1$ it can be determined
numerically.  $g(x)$ and $h(x)$ are shown in Figure~\ref{f02}.

We employ $g(x)$ to characterize the two physicality requirements,
\mbox{$M_r\ge0$} and \mbox{$dM_r/dr\ge0$}, in their difference equation form
(Equations~\ref{eq08} and \ref{eq09}).  In both cases $x_0$ is set to a large
value that we don't expect to exceed.  In the former case we set it to
$10^{18}$ \Msun; in the latter we can use the equivalent of
$d\log(M_r)/d\log(r) \leq 3$, corresponding to the upper limit for a
monotonically decreasing density profile.  We can ignore the normalization and
set $K=1$ since we will be weighting the physicality penalties against the
fidelity penalty.  The complete cost function $f$ for the minimization is thus:
\begin{equation}
f = \sum^{N}_{i=1} \, [ \, A_1 \, g(Y_i) + A_2 \, g(Z_i) \, ] \, + \chi^2
\label{eq14}
\end{equation}

\noindent  
The weights $A_1$ and $A_2$ and the exponent $\eta$ are chosen to determine
whether the departure from the unconstrained profile is reasonable to achieve
a physically acceptable solution.

Conveniently, this technique has a built-in gauge of the fidelity, the $\chi^2$
value of the constrained profiles (although it should be noted that it is not
distributed as the classical $\chi^2$).  If $\chi^2 \lesssim 1$ then we
consider the new profile to be a reasonable excursion within the profile's
uncertainty envelope.  Moreover, we identify the constrained profiles
$(\tilde{\rho},\tilde{T},\tilde{M_r})$ as
{\it a closer approximation to the true profiles}, under the assumption that
our model is correct.  Conversely, $\chi^2 \gg 1$ indicates that the ($\rho,T$)
profiles require excessive alteration to produce a physically sensible mass
profile.  This could indicate that our assumptions of spherical symmetry and/or
hydrostatic equilibrium are invalid, that there is significant spatial or
spectral substructure, or that our thermal emission model is inadequate.
Regardless of the root cause, the accuracy of such a profile is suspect.

\citet{nulsen} (hereafter NB95) developed a non-parametric approach to this
problem which also makes use of the fact that the enclosed gravitating mass of
a cluster must be monotonically increasing.  With their method one obtains a
series of interdependent constraints on the mass at each point in $r$.  These
constraints are translated into a set of likelihood functions which are assumed
to be independent for computational ease.  These likelihood functions are
jointly maximized to derive the mass profile.  Our method is similar in that it
makes use of a likelihood function based upon excursions within an uncertainty
envelope.  The methods differ significantly, however, in the deprojection
algorithm (NB95 use onion-peeling), and in the enforcement of the mass
constraints.  In NB95, the mass constraints are absolute; in our method they
can be invoked to any degree -- they can be rigidly enforced, completely
ignored, or somewhere in between.  The $\chi^2$ metric of the fidelity is a
powerful tool.  If the physicality weight is set to an arbitrarily large value,
we force the profile to monotonically increase, in effect obtaining the result
of the NB95 technique.  As a biproduct we obtain the plasma temperature and
density profiles which are required, and the $\chi^2$ value immediately tells
us how likely it is that the temperature and density measurements conspired to
achieve this condition.  In cases where rigid imposition of the constraints
yields an unacceptably high $\chi^2$ value, one can experiment with different
weights and examine the resulting profiles to ascertain, for example, if the
problem is due to a single outlier, or is more systemic, possibly indicative of
a breakdown of hydrostatic equilibrium.  Thus the flexibility of our method
with regard to the mass constraints allows us to extract information about the
dynamical state of the X-ray plasma and the presence of statistical anomalies
in the data.

\section{Choosing a model:  the $F$ test and Markov Chain Monte Carlo sampling}
\label{mcmc}

In the previous section we assumed that the underlying model is correct.  There
is some uncertainty, however, about the form the model should take.  As
mentioned previously, the best candidates for this type of analysis are relaxed
cooling flow clusters.  These systems are currently not well understood, yet
the way we model the cool plasma can have a significant effect on the resulting
mass profile, and conclusions we may draw about the properties of dark matter
particles or the evolution of large scale structure \citep{abg}.

A standard procedure for choosing between a simple model ${\sf M^s}$ and a
complex model ${\sf M^c}$ is to utilize the $F$ test \citep{bevington}.  This
is done by computing the $F$ value for the data set ${\sf D}$:
\begin{equation}
F = \frac{\chi^2({\sf M^s|D}) -
\chi^2({\sf M^c|D})}{\chi^2({\sf M^s|D})/\nu({\sf M^s})}
\label{eq15}
\end{equation}

\noindent and comparing it to the standard $F$ distribution.  Here
$\chi^2({\sf M^s|D})$ and $\chi^2({\sf M^c|D})$ are the sum of the squares of
the error-weighted residuals in the spectroscopic least-squares fit to the
simple and complex models, respectively, and $\nu({\sf M^s})$ is the number of
degrees of freedom in the simple model.

Unfortunately, the standard $F$ test is not applicable in this context. As
\citet{protassov} have pointed out, the standard $F$ test is valid only in
cases where the simple model is nested within the complex model.  In the
present case, the simple model lies on a boundary of the complex model.  That
is, the simple model is a special case of the complex model, with the
normalization of the second emission component in the core set to zero.  This
means that its $F$ distribution may deviate significantly from the norm (see
Figure~\ref{f03}).  Instead we must construct an {\it empirical} $F$
distribution by sampling the probability distribution function of the simple
model, simulating a data set for each sample item, and applying both models
to the simulated data.  Once our $F$ distribution has been constructed, we can
judge the significance of the extra emission component based upon the location
within the distribution of the $F$ value for the data \citep{protassov}.

We employ the Markov Chain Monte Carlo (MCMC) sampling technique
\citet{neal,vandyk,lewisbridle,hobson} to build a large sample of data
realizations from which to construct an empirical $F$ distribution.  Let
$P(\bf x)$ represent the posterior probability distribution function of the
parameters ${\bf x} = x_1, x_2, ..., x_N$ determined by fitting the simple
model ${\sf M^s}$ to a real data set ${\sf D}_0$.  We can sample $P({\bf x})$
by taking a rejection-based random walk through the parameter space.  We define
a transition probability $T({\bf x}_n,{\bf x}_{n+1})$ as the probability of
moving from an initial set of parameters ${\bf x}_n$ to a new set of parameters
${\bf x}_{n+1}$.  $T$ depends on the value of the posterior distribution at the
original and new parameter sets.  Let $q({\bf x}_n,{\bf x}_{n+1})$ be an
arbitrary proposal distribution; that is, the probability that the new proposed
parameter set is ${\bf x}_{n+1}$ given that we are presently at ${\bf x}_n$.
If we accept the proposed parameter set with probability $\alpha$, then
\begin{equation}
\alpha({\bf x}_{n},{\bf x}_{n+1}) =
\frac{T({\bf x}_n,{\bf x}_{n+1})}{q({\bf x}_n,{\bf x}_{n+1})}
\label{eq16}
\end{equation}

\noindent which takes into account the odds of actually stepping to the new
location in parameter space, $q({\bf x}_n,{\bf x}_{n+1})$, and the odds of such
a transition between the two locations being accepted,
$T({\bf x}_n,{\bf x}_{n+1})$.  The acceptance probability $\alpha$ is
calculated from
\begin{equation}
\alpha({\bf x}_{n},{\bf x}_{n+1}) = {\rm min}\left[ 1 , \,
\frac{P({\bf x}_{n+1}) \, q({\bf x}_{n+1},{\bf x}_n)}
{P({\bf x}_n) \, q({\bf x}_n,{\bf x}_{n+1})} \right]
\label{eq17}
\end{equation}

In our case we use a particular form of MCMC sampling called the Metroplis
algorithm (see, e.g., \citet{neal}) which uses a symmetric proposal
distribution function:
\begin{equation}
q({\bf x}_n,{\bf x}_{n+1}) = q({\bf x}_{n+1},{\bf x}_n)
\label{eq18}
\end{equation}

This prescription for wandering through parameter space constitutes a Markov
chain since each new parameter set is chosen according to a probability
distribution function that depends only upon the previous set of values.
In the case of the Metropolis algorithm, it is straightforward to show that
$P({\bf x})$ is an invariant distribution of the Markov chain.  Using
Equations~\ref{eq16} and \ref{eq17} we have
\begin{eqnarray}
P({\bf x}_n) \, \, T({\bf x}_n,{\bf x}_{n+1}) & = & P({\bf x}_n) \, \,
q({\bf x}_n,{\bf x}_{n+1}) \, \, \alpha({\bf x}_n,{\bf x}_{n+1}) \nonumber \\
 & = &
q({\bf x}_n,{\bf x}_{n+1}) \, \, {\rm min}[ P({\bf x}_n), P({\bf x}_{n+1}) ]
\label{eq19}
\end{eqnarray}

\noindent Making use of the symmetry of $q$ in the Metropolis algorithm, we
have
\begin{eqnarray}
\hspace{90pt} & = & q({\bf x}_{n+1},{\bf x}_n) \, \,
{\rm min}[ P({\bf x}_n), P({\bf x}_{n+1}) ] \nonumber \\
 & = & P({\bf x}_{n+1}) \, \, q({\bf x}_{n+1},{\bf x}_n) \, \,
\alpha({\bf x}_{n+1},{\bf x}_n) \nonumber
\end{eqnarray}
\noindent resulting in
\begin{equation}
P({\bf x}_n) \, \, T({\bf x}_n,{\bf x}_{n+1}) =
P({\bf x}_{n+1}) \, \, T({\bf x}_{n+1},{\bf x}_n)
\label{eq20}
\end{equation}

\noindent This statement of detailed balance demonstrates that $P({\bf x})$ is
a stationary distribution of the Markov chain.  This is necessary, though not
sufficient, to ensure that we can sample $P({\bf x})$ directly using an
appropriately selected chain of Monte Carlo simulations.  The other necessary
condition, ergodicity, ensures that any substring of the Markov chain
will asymptotically approach $P({\bf x})$ regardless of the initial conditions,
although a derivation of this property is beyond the scope of this paper.  For
a complete discussion see \citet{neal}.

In many applications of MCMC sampling one pays special attention to the finite
``burn-in'' period during which the Markov chain equilibrates.  The length of
the burn-in phase depends upon the sensibility of the starting point, and the
appropriateness of the scale chosen for the proposal probability distribution
step.  This is not a consideration in our case because we start each MCMC
sample at the (already known) peak of the probability distribution function
$P({\bf x})$.

Given $P({\bf x})$ computed by the MCMC process, we can construct an empirical
$F$ distribution and perform an $F$ test on the significance of a second
emission component in the core.  The entire procedure is as follows:

\begin{verse}

1. Model the real data set ${\sf D}_0$ with $\sf M^s$; call the best-fit
parameters ${\bf x}_0^{\sf s}$.

2. Use XSPEC to calculate $P({\sf D}_0|{\bf x})$ (i.e., the likelihood).

3. Use Bayes' Theorem to calculate $P({\bf x}|{\sf D}_0)$ (see
Equation~\ref{eq07}).  We discard all unphysical excursions in parameter
space, i.e. where $T<0$ or $\rho<0$.  (In practice, we discard at the level of
the model normalization, not $\rho$, but $\rho$ is simply a function of the
normalization and the binning geometry.)

4. Create a large sample of model parameters ${\bf x}_i^{\sf s}$ using
$P({\bf x}|{\sf D}_0)$ and the Metropolis algorithm form of the MCMC technique.
For each ${\bf x}_i^{\sf s}$, compute a fake data set ${\sf D}_i$, including
instrumental effects of the {\it Chandra} telescope and detectors, as well as
counting statistics.

5. Model each ${\sf D}_i$ using both $\sf M^s$ and $\sf M^c$.

6. For each pair of models tabulate its $F$ value given by Equation~\ref{eq15}.

7. Bin up the set of $F$ values, creating an unnormalized histogram , and
superimpose the $F$ value of the original data.

\end{verse}

In practice this recipe is computationally intensive, not because of any
features of the MCMC sampling {\it per se}, but because each of the faked
spectra must be modelled twice.  For a sample size of 1000 simulations,
XSPEC must simulate 1000 spectra and calculate 2002 sets of best-fit
values for the model parameters (including the original data).  This fact
leads us to simplify the method.  First, in order to reduce the modelling
time, we have adopted a simplified core-halo geometry.  In this scheme the
``core'' is represented by a single shell (in this case a sphere), while the
halo is represented by another shell.  Thus $\sf M^s$ contains four parameters,
the temperature and density of each of the two shells, while $\sf M^c$ contains
six, the additional two parameters representing the temperature and density of
a second cospatial emission component in the core.  This simplification also
greatly improves the numerical stability of the fitting procedure.  The
algorithm (steps 1-6 above) is implemented in a Tcl script run within XSPEC.

Once we have completed step 7 we can distinguish between the models.  The
location of the $F$ value of the data within this empirical $F$ distribution
contains information regarding the relative merit of $\sf M^s$ and $\sf M^c$.
We define the significance $S$ of the distribution as
\begin{equation}
\label{eq21}
S = \frac{\int_{0}^{F_{data}} N(F) \, dF}{\int_{0}^{\infty} N(F) \, dF}
\end{equation}

\noindent The signficance $S = 1 - P_f$, where $P_f$ is the probability that
the simple model constitutes the better description, and that the $F$ value of
the data is this large strictly by chance.  Thus, for a one-parameter model,
$S = 0.68$, 0.90, and 0.99 may be interpreted as 1-, 2- and 3-$\sigma$
detections of the additional component.  We checked the sensitivity of this
method by applying it to five simulated data sets with known mixtures of hot
($T=5.0$ kev) and cold ($T=1.0$ kev) X-ray plasmas in pressure equilbrium.
We use mass ratios of $M_C/M_H = 0.000, 0.0222, 0.0500, 0.0857$, and 0.133,
and run 100 MCMC simulations for each.  Figure~\ref{f04} shows an empirical
$F$ distribution for each of these cases.  The $F$ value of the data is shown
with a dashed vertical line.  For a multiphase plasma of which only 2.2\% is
in the cold component, the detection of the multiphase plasma is better than
2$\sigma$.  At 5\% and greater, the detection is statistically highly
significant.

\section{Application to Chandra Clusters}
\label{app}

We illustrate these techniques using a X-ray observations of a sample of
bright, apparently relaxed galaxy clusters (see Table~\ref{t01}).  Each of
these clusters (except A1689) contains a significant amount of cooler plasma in
its core, and is known as a classical ``cooling flow cluster'' since the core
radiative cooling time is shorter than the age of the cluster.  
We prepared each archived data set as described in \citet{abg} and modelled the
emission using the two models described in \S\ref{depro}.  In the first model,
each shell contains isothermal plasma ($N_c=0$), while in the other, the
central two shells are also allowed a second (cooler) emission component
($N_c=2$), and the best-fit parameter values are obtained iteratively using
XSPEC.  Figures~\ref{f05}-\ref{f14} show baryon density, baryon temperature,
enclosed spherical gravitating mass, and enclosed cylindrical gravitating mass
profiles for each cluster in the sample.  The left panels show $N_c=0$ models;
the right show $N_c=2$ model.  The unconstrained profiles are shown as data
points, with red [blue] symbols representing the hot [cool] plasma components.
(Data points which lie outside the ordinate range are indicated by arrows.)
Constrained profiles are drawn as solid curves; the density and temperature
contributions to the $\chi^2$ value of the constrained solutions are listed in
Table~\ref{t02}.  The last column in that table lists the MCMC significance $S$
of the presence of multiphase gas in each cluster core as determined from the
MCMC-derived empirical $F$ distributions shown in Figure~\ref{f15}.  For the
subset of five clusters which are arguably the most relaxed, and are best
described by an NFW profile (A1689, A1835, A2029, MS1358 and MS2137; see AB),
we overlay the projected mass profiles with weak and/or strong lensing
measurements from the literature (see Table~\ref{t03} for references).  Weak
lensing mass profiles or isothermal sphere fits to weak lensing data are shown
in violet; strong lensing measurements are shown in green.  We adopt $H_0=67$
km s$^{-1}$ Mpc$^{-1}$, $\Omega_\Lambda=0.7$ and $\Omega_m=0.3$.

A detailed analysis of these profiles and their consequences for cosmology and
dark matter candidates is in preparation (AB), so we make only a few brief
comments here.  Of the ten clusters presented here, only HydraA did not admit
a second emission component in the core -- attempts to add one resulted in the
temperature of the second component being set to the temperature of the first
in the fitting procedure.  This is consistent with previous {\it Chandra}
\citep{david} and {\it XMM-Newton} \citep{kaastra} results.  Four of the
clusters show evidence for multiphase core plasma at the 99\% significance
level -- A2029, A2204, MS1358 and ZW3146 (see Table~\ref{t02}).  This is
consistent with the study of \citet{kaastra}, which finds evidence for
multiphase plasma in many clusters.  (Note that, with 1000 MCMC simulations per
cluster, the precision of the significance estimate is $\sim$3\%.)  For
comparison, a cluster which contained no second plasma component would, on the
average, show an MCMC significance of order 0.5, or 50\%.  These clusters show
a greater difference in their core masses between the uni- and multiphase
models.  This phenomenon is illustrated for two clusters, MS1358 and MS2137, in
Figures~\ref{f16} and \ref{f17}.  The mass of MS1358 in the multiphase model is
about a factor of two larger in than the uniphase mass, and its MCMC
significance $S$=0.987.  MS2137, on the other hand, with an MCMC significance
of 0.271, shows very little difference between the two mass models.  This is
perhaps unsurprising, as one would expect that a significant amount of
cospatial cool plasma in its core would not only display a clear observational
signature but would also affect the equilibrium configuration of the plasma.
MS2137 and ZW3146 also provide an interesting contrast.  In the
pre-{\it Chandra/XMM-Newton} era, both clusters were reported to harbor cooling
flows with mass deposition rates in excess of 1000 \Msun y$^{-1}$
\citep{allenII}, yet they show remarkably disparate evidence for multiphase
plasma in their cores ($S$(MS2137)=0.271; $S$(ZW3146)=0.989).  This suggests
that there may be more than one mechanism at work responsible for the presence
of  1 keV plasma at the center of galaxy clusters.

For each constrained reconstruction we use $\eta=2.5$ and physicality weights
$A_1=A_2=A_{12}$ originally set to $1\tenup{-6}$.  While this often will not
rigidly enforce constraint equation~\ref{eq05}, it is usually sufficient to
enforce equation~\ref{eq06}.  In those cases where not even equation~\ref{eq06}
is satisfied, we increased $A_{12}$ by factors of 10 until the resulting
profile was nonnegative everywhere if possible.  This is the origin of the
unacceptably high values of the $\chi^2$ fidelity measures for the $N_c=2$
model of A2104 and the $N_c=1$ model of HydraA.  Obviously, the fidelity of a
constrained profile tends to be greater when the unconstrained profile shows
only one or two significant outliers.  In some cases (e.g.\ A1795) a very
small adjustment to the temperature profile produces a large change in the
derived mass.  This is due to the competition between the derivatives in
equation~\ref{eq02} -- if a statistical fluctuation in the temperature
measurement is large enough (and positive) it will swamp the surface brightness
decrement at that radius, resulting in a very low or even negative mass.  In
cases where $d\log T/d\log r \gtrsim -d\log\rho/d\log r$, a relatively minor
adjustment in the temperature can remove an unphysical point from the mass
profile.

The five very relaxed clusters generally show better agreement with weak
lensing mass measurements than they do with strong lensing.  The reprojected
profile A1689 shows the least agreement with the weak lensing, differing by
up to a factor of two at some radii, although the lensing profile is a singular
isothermal sphere (IS) fit to weak lensing measurements \citep{kcs}.  The
strong lensing measurement of \citet{wu}, however, exceeds our profile again by
a factor of 2.  The error bars are derived by assuming two values of the
(unknown) strong lensing arc redshift (0.8 and 2.0); the discrepancy is thus
unlikely to be due to incorrect source redshift.  Our profile is also
systematically in excess of the IS fit to weak lensing observations of A1835
\citep{cs}, although it is consistent with the strong lensing point of
\citet{allen}.  A2029 and MS1358 show remarkable agreement with weak lensing
data \citep{menard,hoekstra}, although the strong lensing measurement in
MS1358 \citep{franx,allen} is moderately discrepant, as is the strong lensing
measurement of MS2137 \citep{sand}.  These issues will be addressed in greater
detail in AB.

\section{Summary}
\label{summary}
We have presented a technique for calculating the dark matter profile of a
spherical, relaxed galaxy cluster.  We have formulated a technique for coping
with statistical uncertainty in the measurement of the cluster plasma
temperature.  We have also described a method for determining whether there is
a statistically significant presence of multiphase plasma in the galaxy cluster
core.  We have applied these tools to a sample of relaxed galaxy clusters
observed with {\it Chandra}, and find that 4/10 require a multiphase treatment
of their core plasma.  Our masses are in broad agreement with weak lensing
studies, though are often exceeded by those derived from strong lensing models.

JSA would like to thank Steve Allen, Aaron Lewis, Jimmy Irwin and Renato Dupke
for many enlightening discussions.  The authors would like to thank the
anonymous referee for comments which have improved this paper.  This work was
supported by SAO Grant AR3-4016X.





\plotone{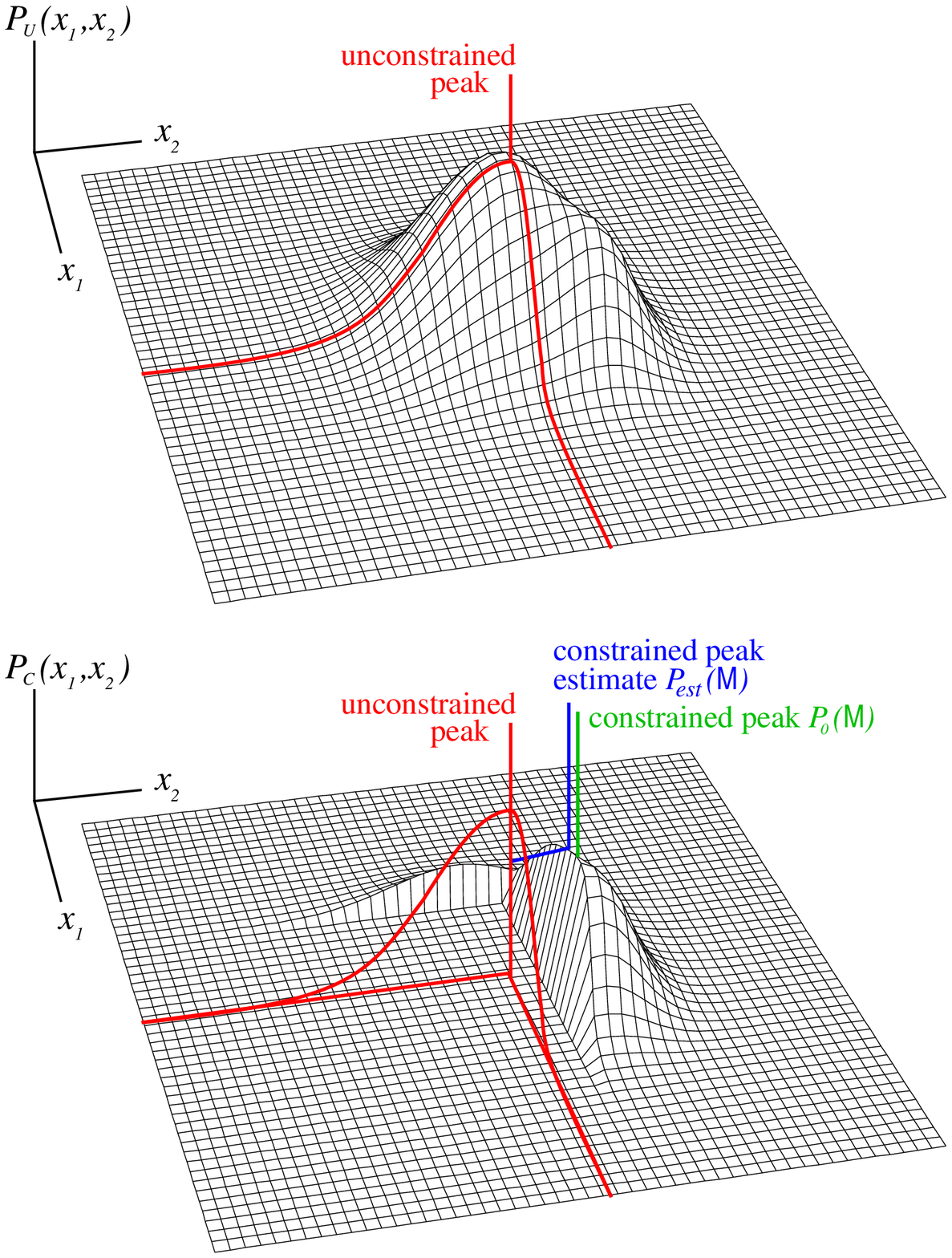}

\figcaption{Schematic representations of the unconstrained probability
distribution function ($P_U$, top), and the constrained distribution ($P_C$,
bottom) for a two-dimensional parameter space.  The estimated peak of the
constrained distribution is shown in blue; the true peak is shown in violet.
(For clarity we have not renormalized $P_C$ after removing the physically
disallowed region of parameter space.)
\label{f01}}


\clearpage
\plotone{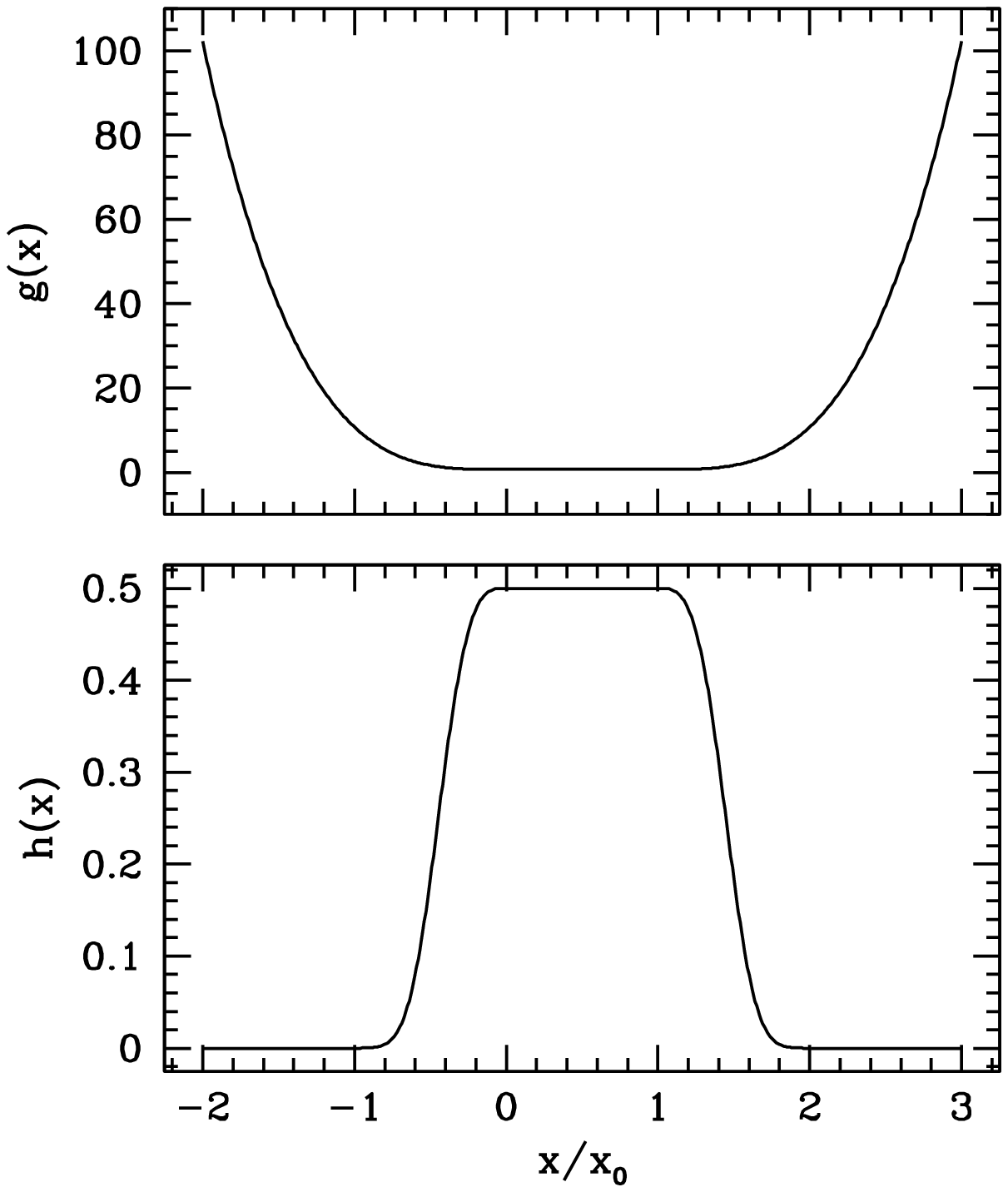}

\figcaption{The penalty function $g(x)$ and the associated (approximate)
probability distribution function $h(x)=e^{-g(x)}$ for $\eta=7/2$.  Here $h(x)$
nearly, but not precisely, normalized; we have set $C=1$ (see
Equation~\ref{eq13}).
\label{f02}}


\clearpage
\plotone{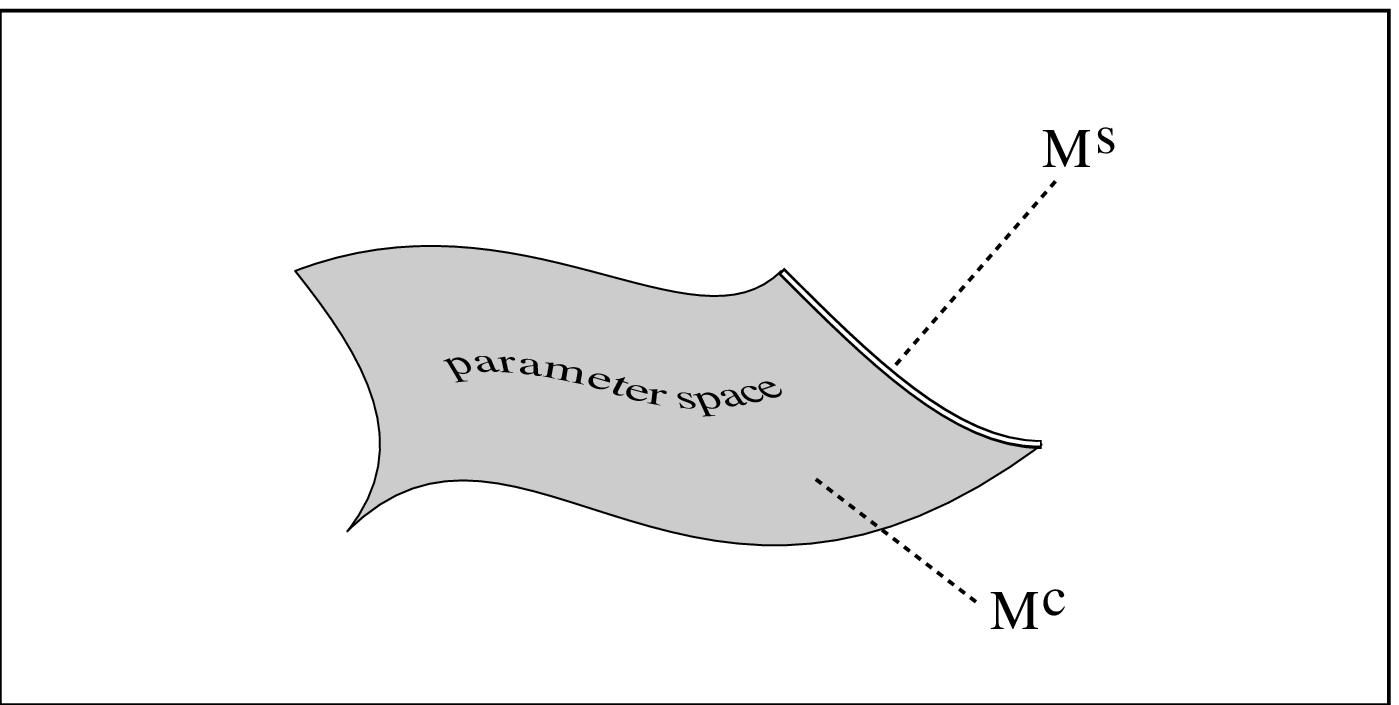}

\figcaption{The non-nested relationship between the simple and complex models.
The simple model corresponds to the complex model with one of the
normalizations set to 0.
\label{f03}}


\clearpage
\plotone{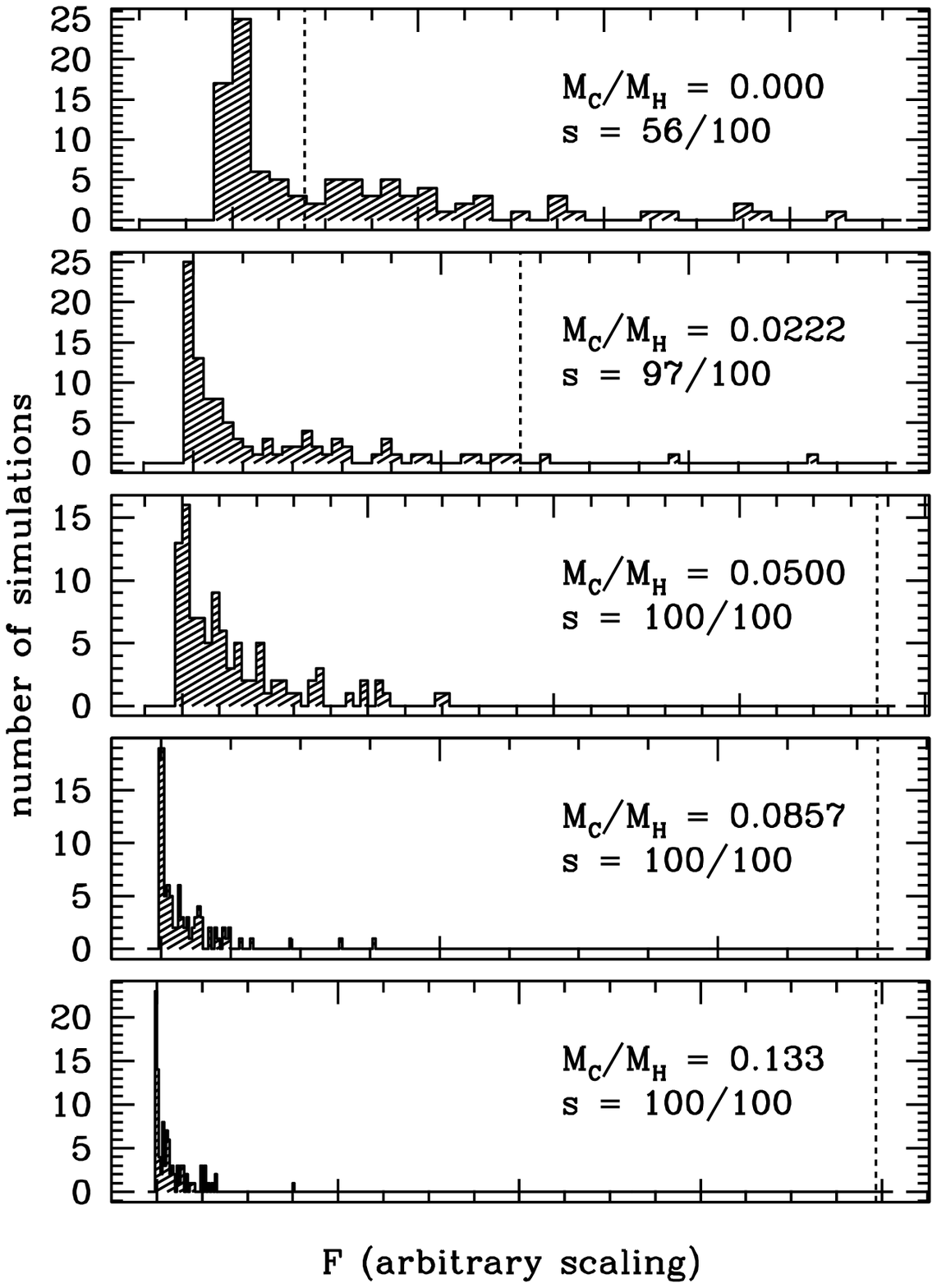}

\figcaption{Empirical $F$ distributions for models ${\sf M^s}$ and ${\sf M^c}$
of five simulated data sets.  The cold to hot plasma mass ratio, assuming
that the two are in pressure equilibrium, is shown for each case, along with
the significance of the presence of the cool component as determined through
100 MCMC simulations (see Equation~\ref{eq21}).  The $F$ value of the original
data set is indicated by a vertical dashed line.
\label{f04}}


\clearpage
\plotone{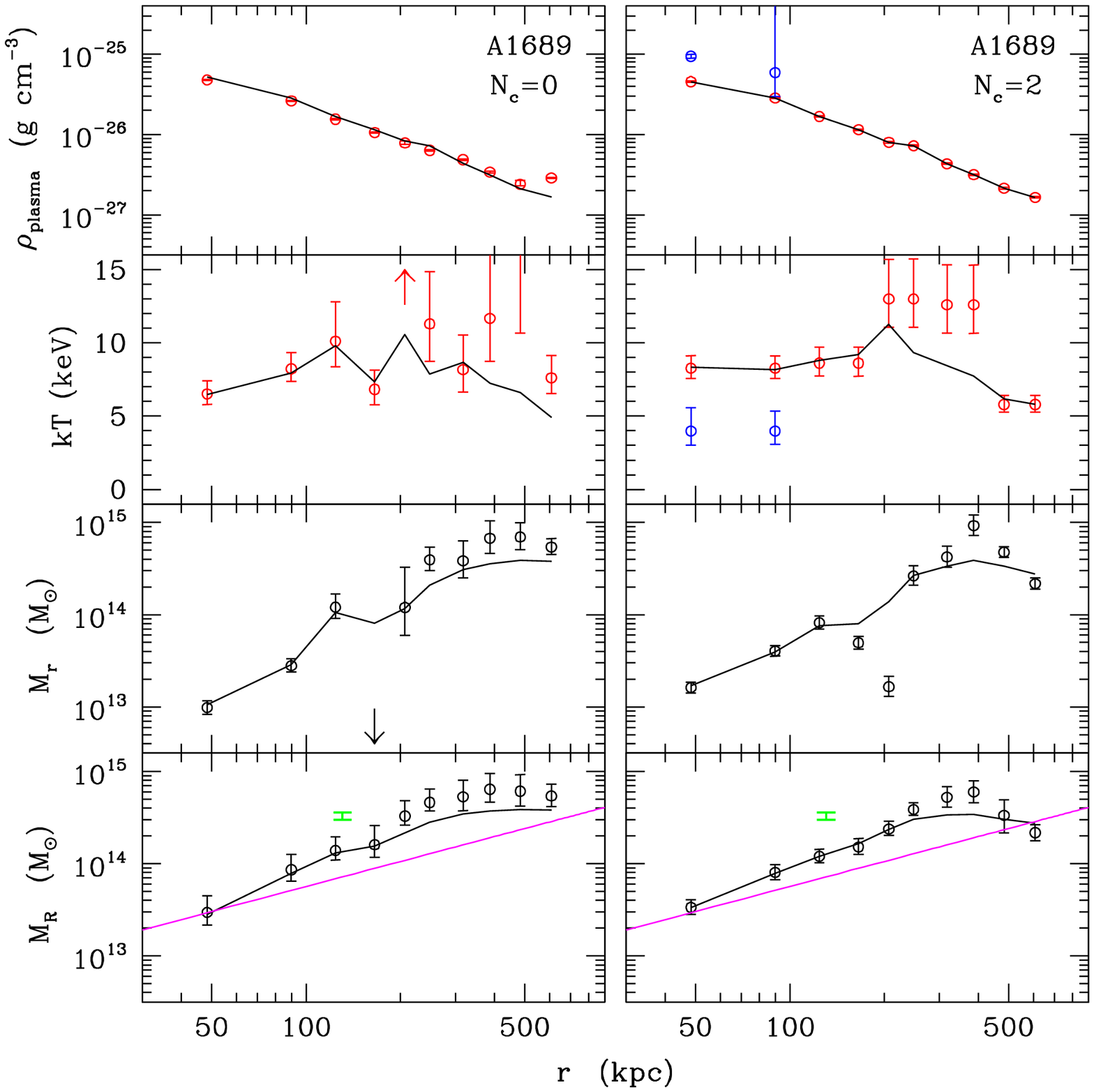}

\figcaption{Baryon density, baryon temperature, spherically enclosed mass, and
cylindrically enclosed (projected) mass profiles of A1689, for $N_c=0$ (left
panels) and $N_c=2$ (right panels).  The hot plasma is shown in red, the cool
component in blue.  Arrows represent points which lie outside the ordinate
range.  Reconstructions adhering to constraint equation \ref{eq06} are shown as
solid black curves.  Weak (violet) and strong (green) gravitational lensing
measurements are shown for comparison in the bottom panels.  (A solid violet
line represents an isothermal sphere fit to the weak lensing data set.)  See
Table~\ref{t03} for lensing references.
\label{f05}}


\clearpage
\plotone{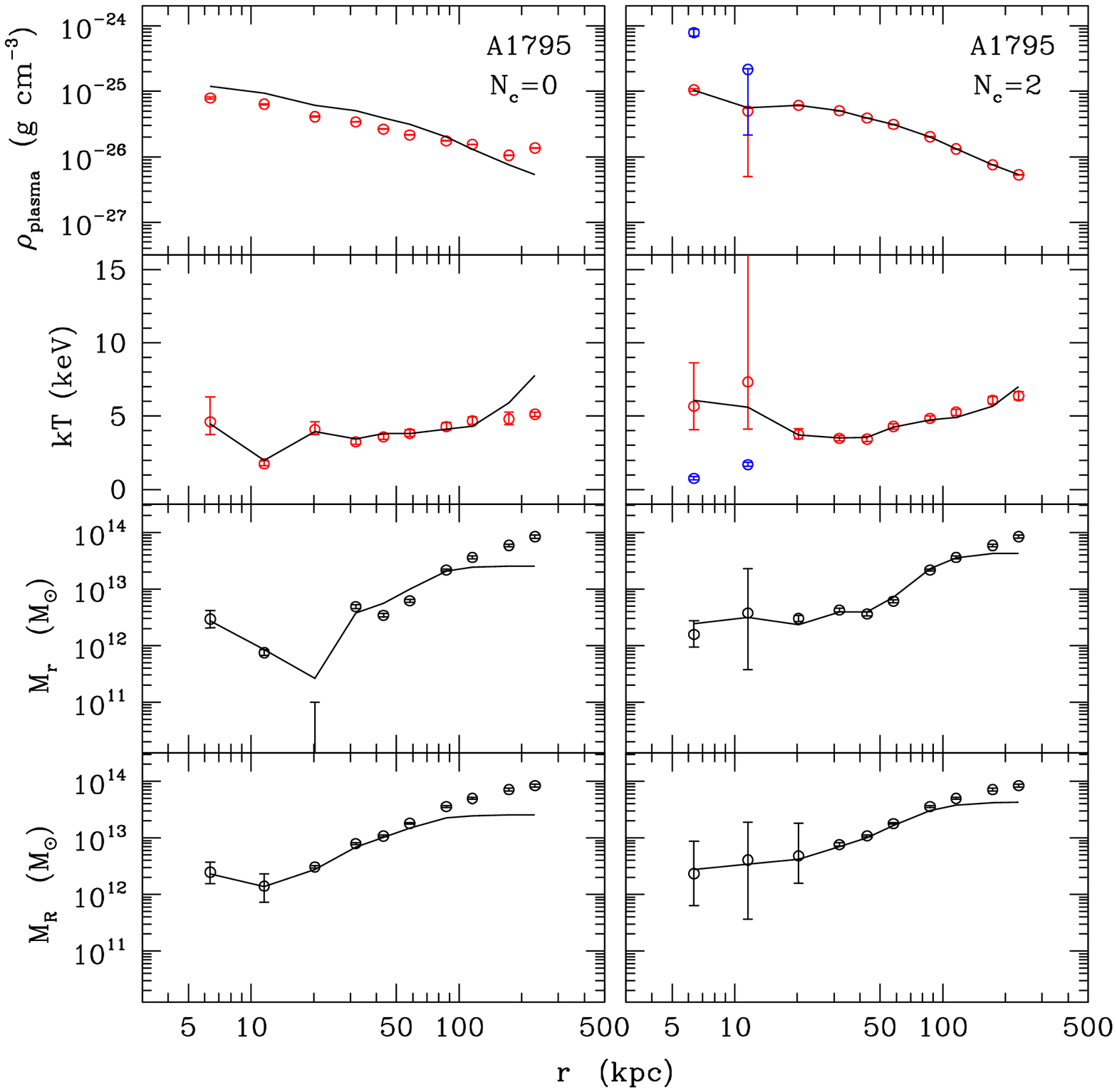}

\figcaption{Same as Figure~\ref{f05}, for A1795.
\label{f06}}


\clearpage
\plotone{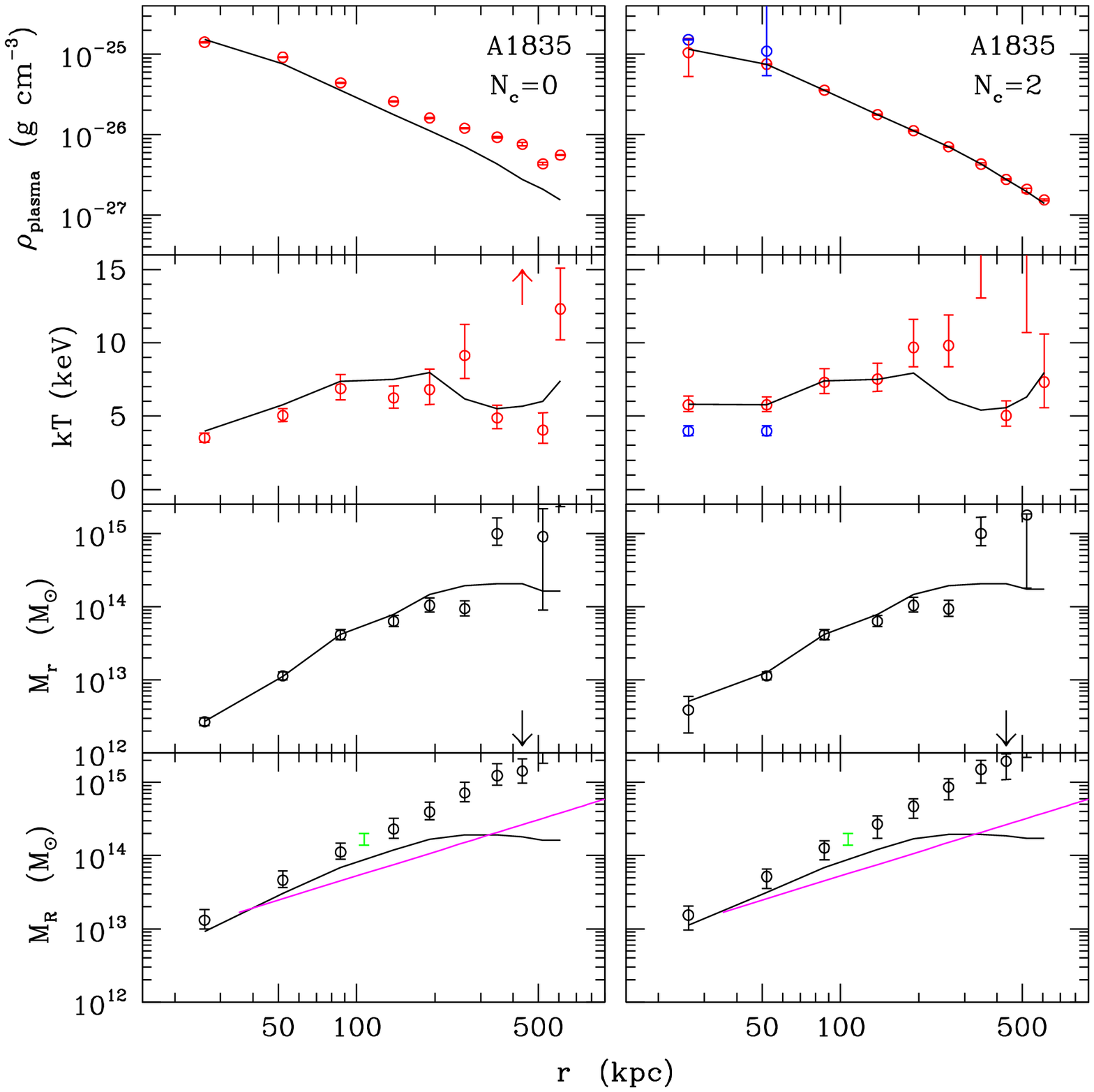}

\figcaption{Same as Figure~\ref{f05}, for A1835.
\label{f07}}


\clearpage
\plotone{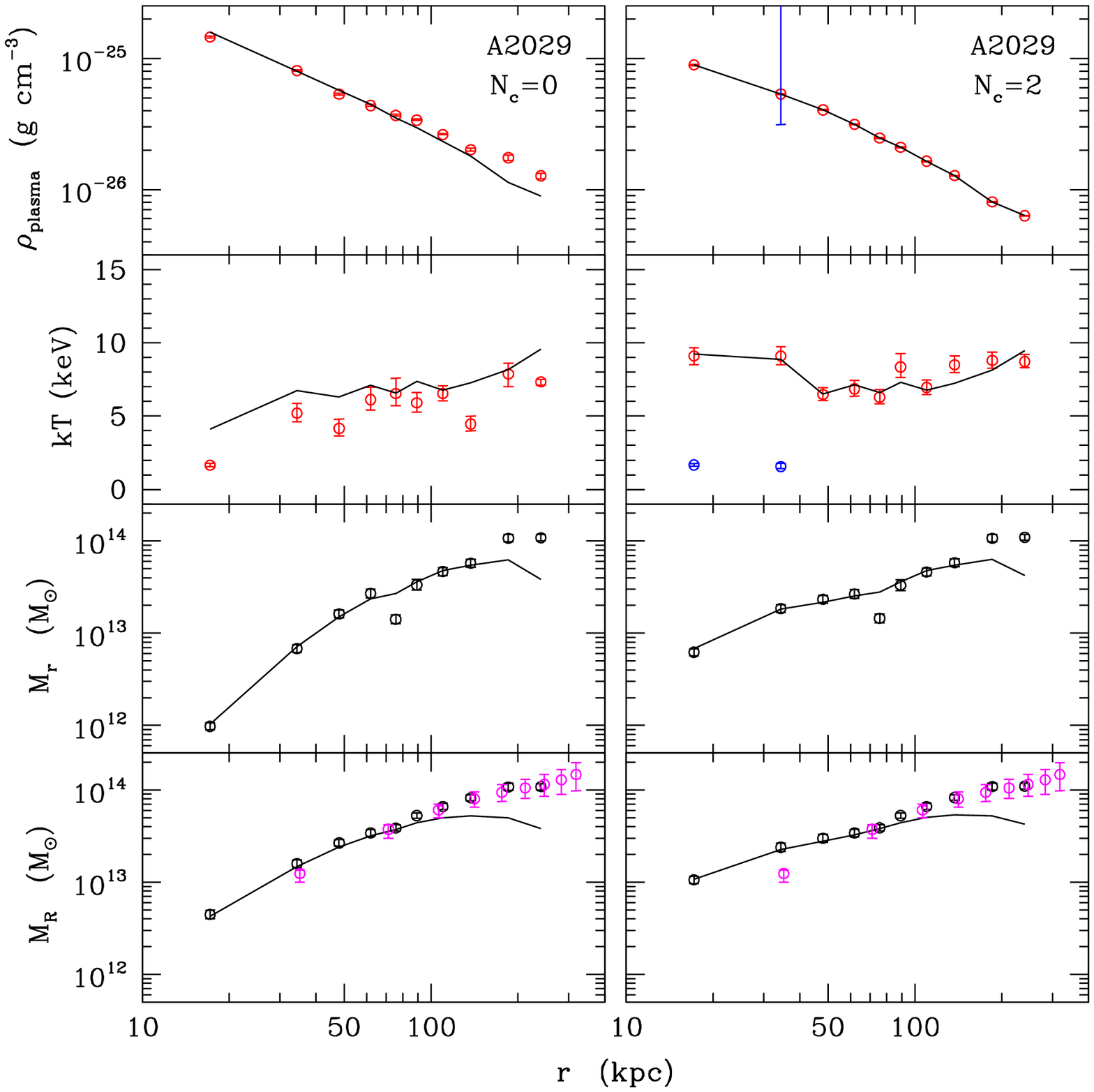}

\figcaption{Same as Figure~\ref{f05}, for A2029.
\label{f08}}


\clearpage
\plotone{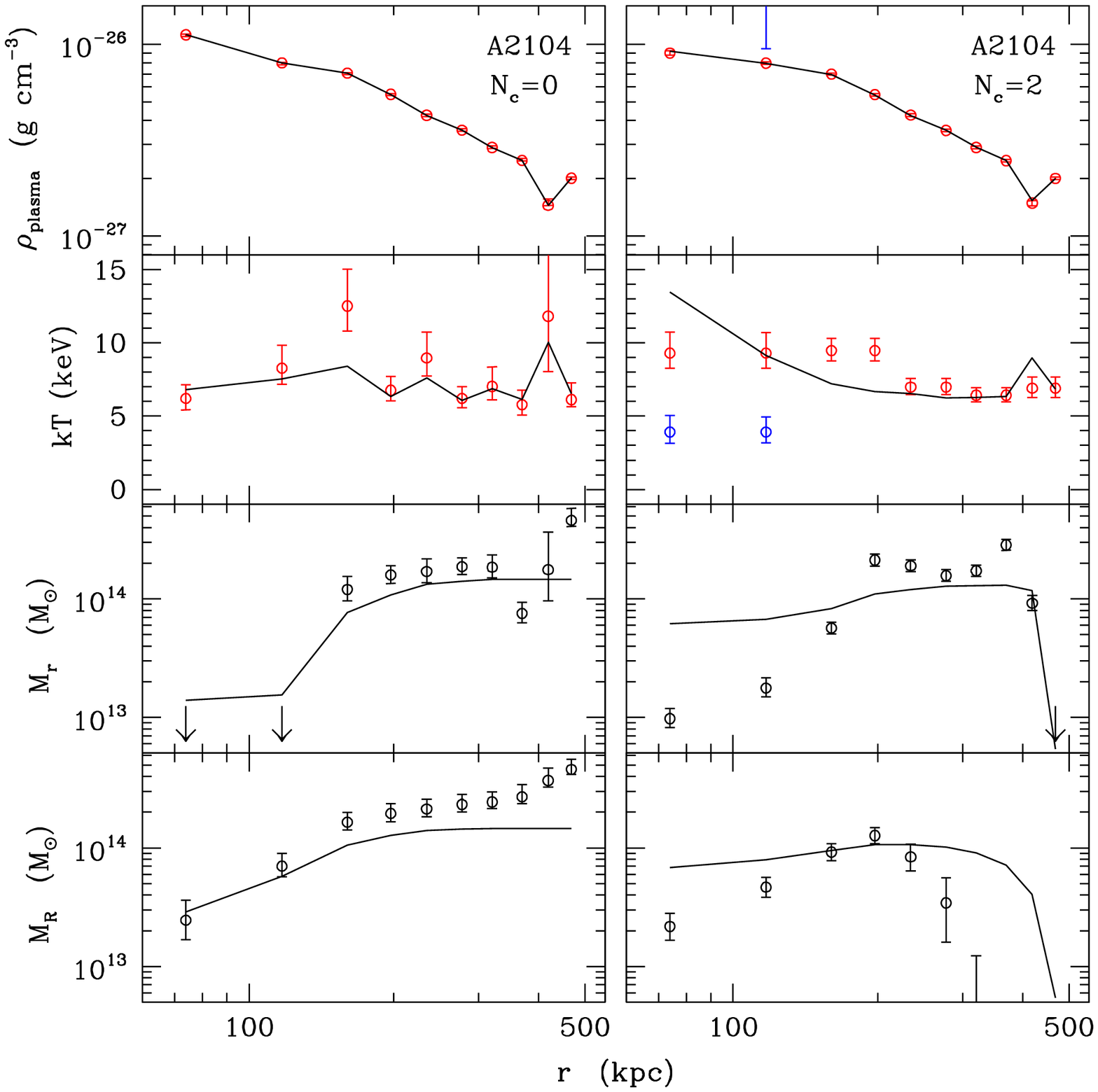}

\figcaption{Same as Figure~\ref{f05}, for A2104.  Note that there is no
reconstruction solution in the $N_c=2$ case which is consistent with constraint
equations~\ref{eq05} and \ref{eq06}.
\label{f09}}


\clearpage
\plotone{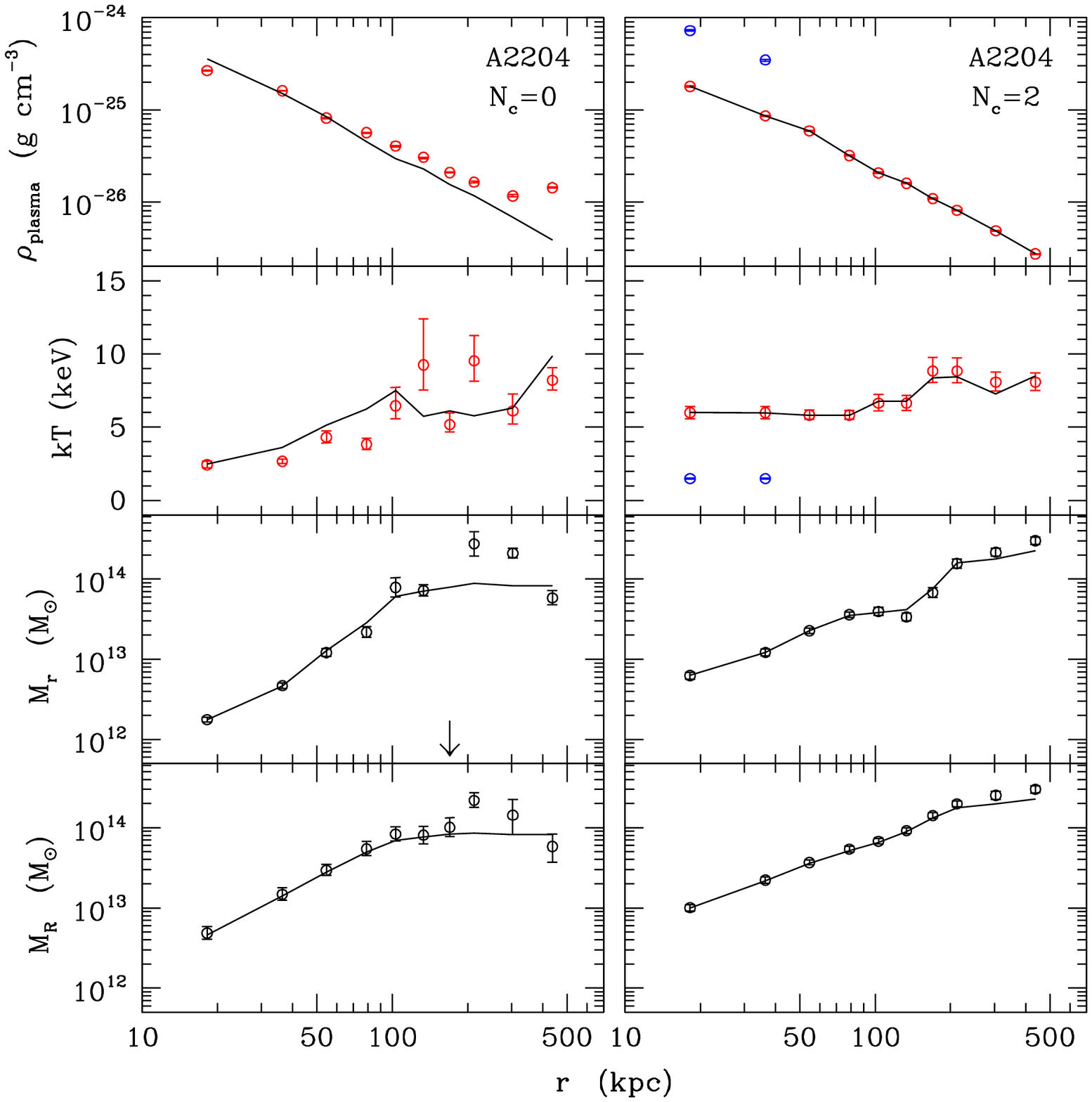}

\figcaption{Same as Figure~\ref{f05}, for A2204.
\label{f10}}


\clearpage
\plotone{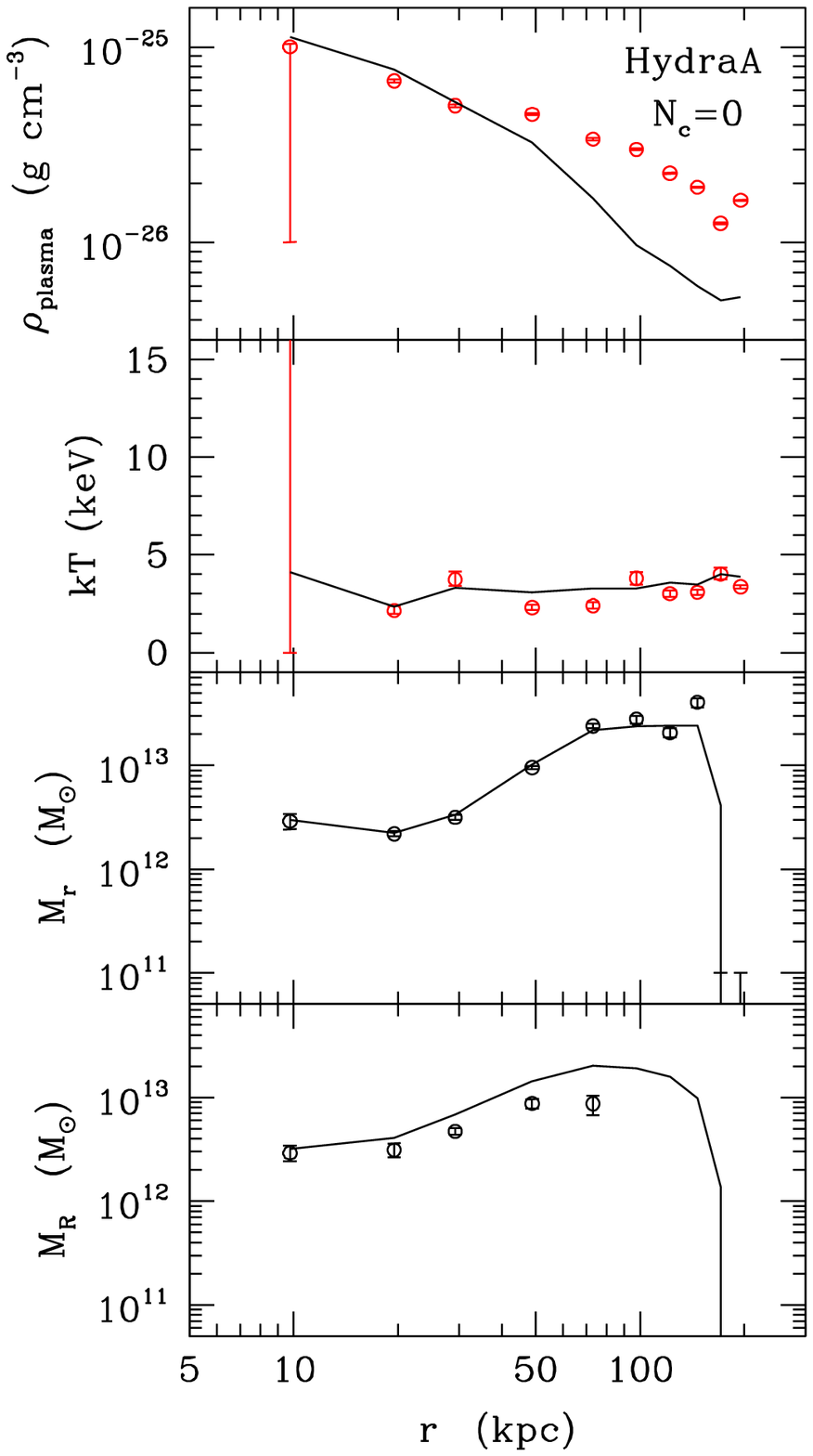}

\figcaption{Same as Figure~\ref{f05}, for HydraA.  Note that there is no
reconstruction solution in the $N_c=1$ case which is consistent with constraint
equations~\ref{eq05} and \ref{eq06}, and that the Hydra A data set did not
admit $N_c=2$ emission models.
\label{f11}}


\clearpage
\plotone{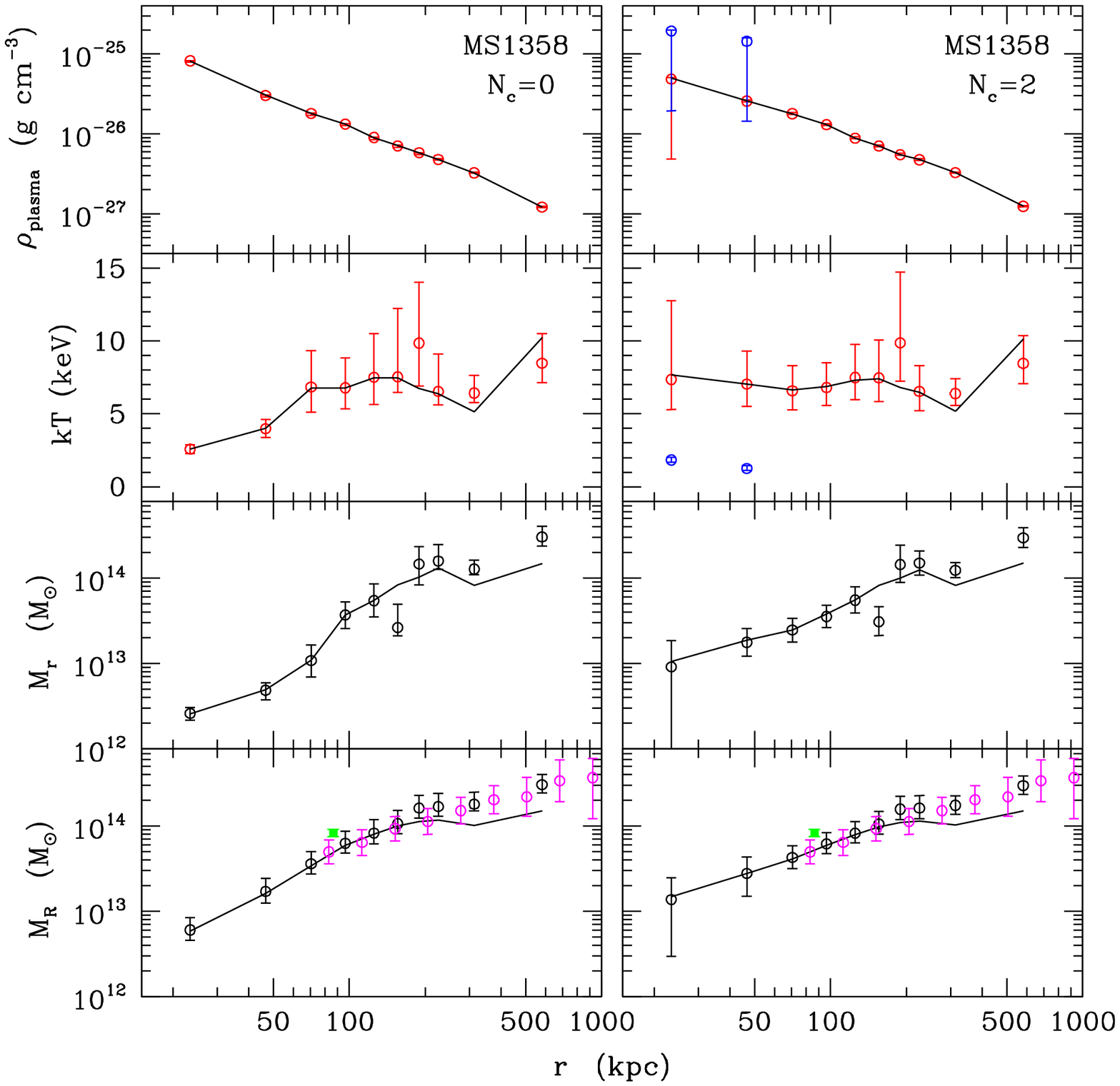}

\figcaption{Same as Figure~\ref{f05}, for MS1358.
\label{f12}}


\clearpage
\plotone{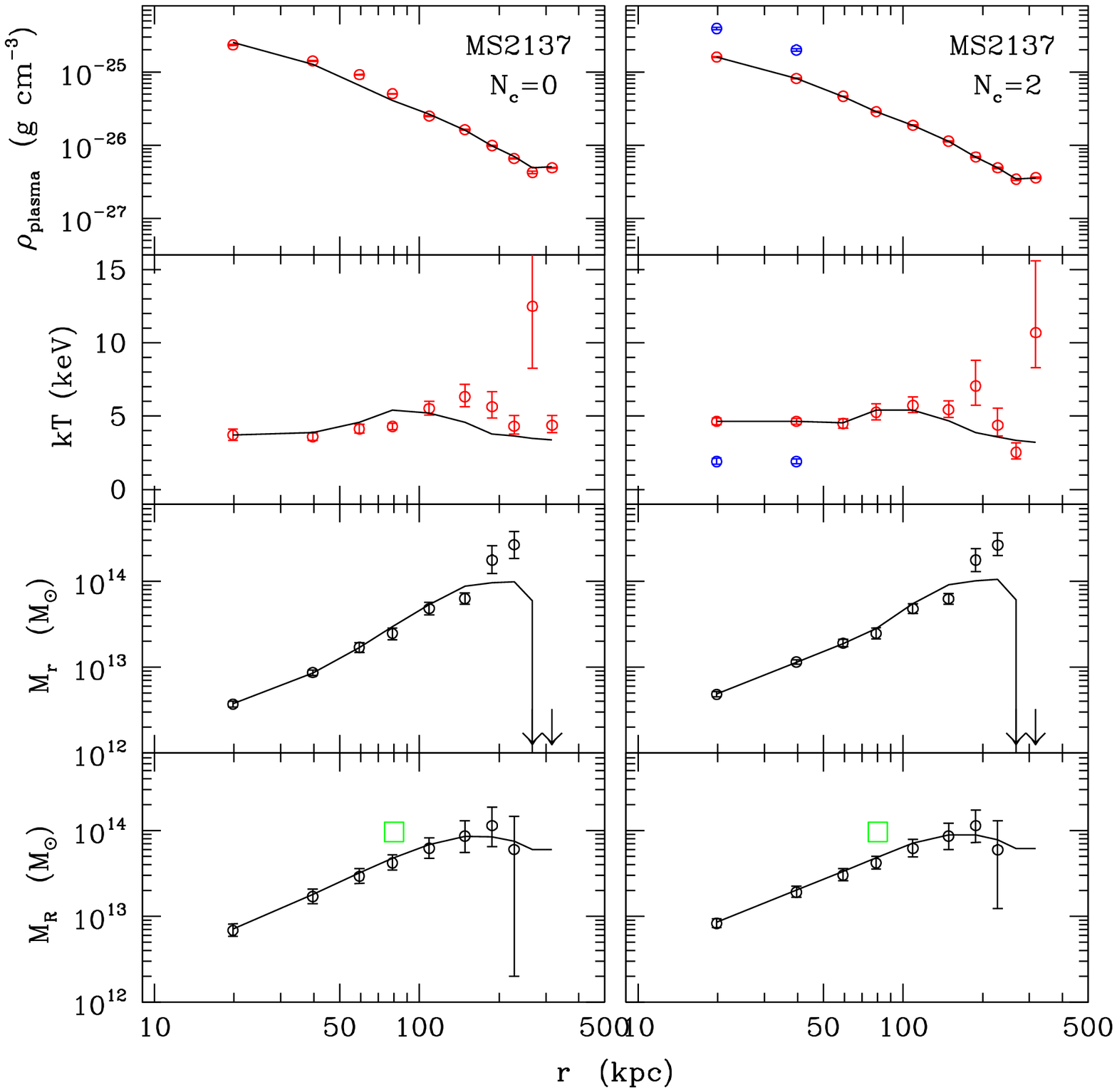}

\figcaption{Same as Figure~\ref{f05}, for MS2137.
\label{f13}}


\clearpage
\plotone{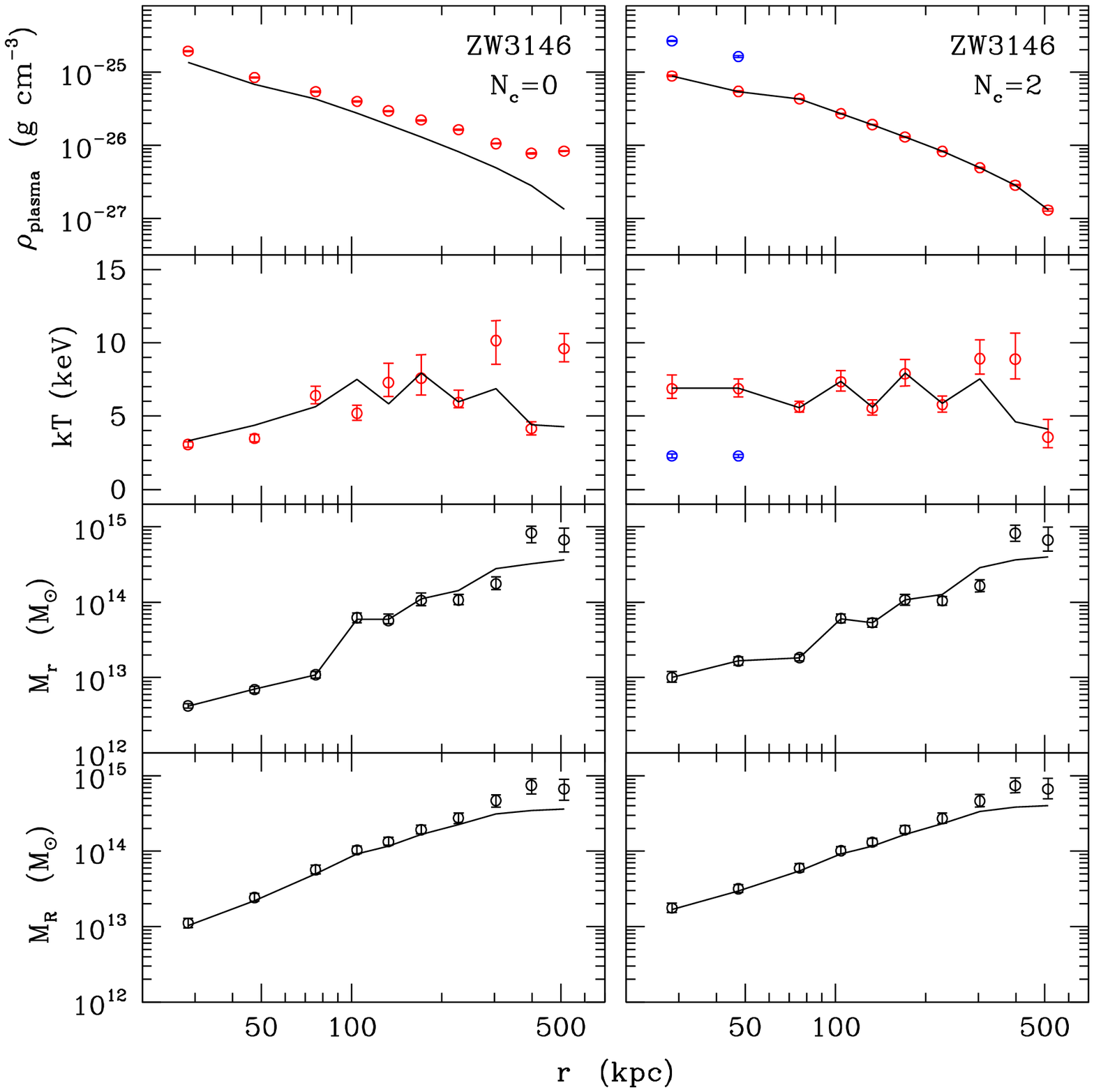}

\figcaption{Same as Figure~\ref{f05}, for ZW3146.
\label{f14}}


\clearpage
\plotone{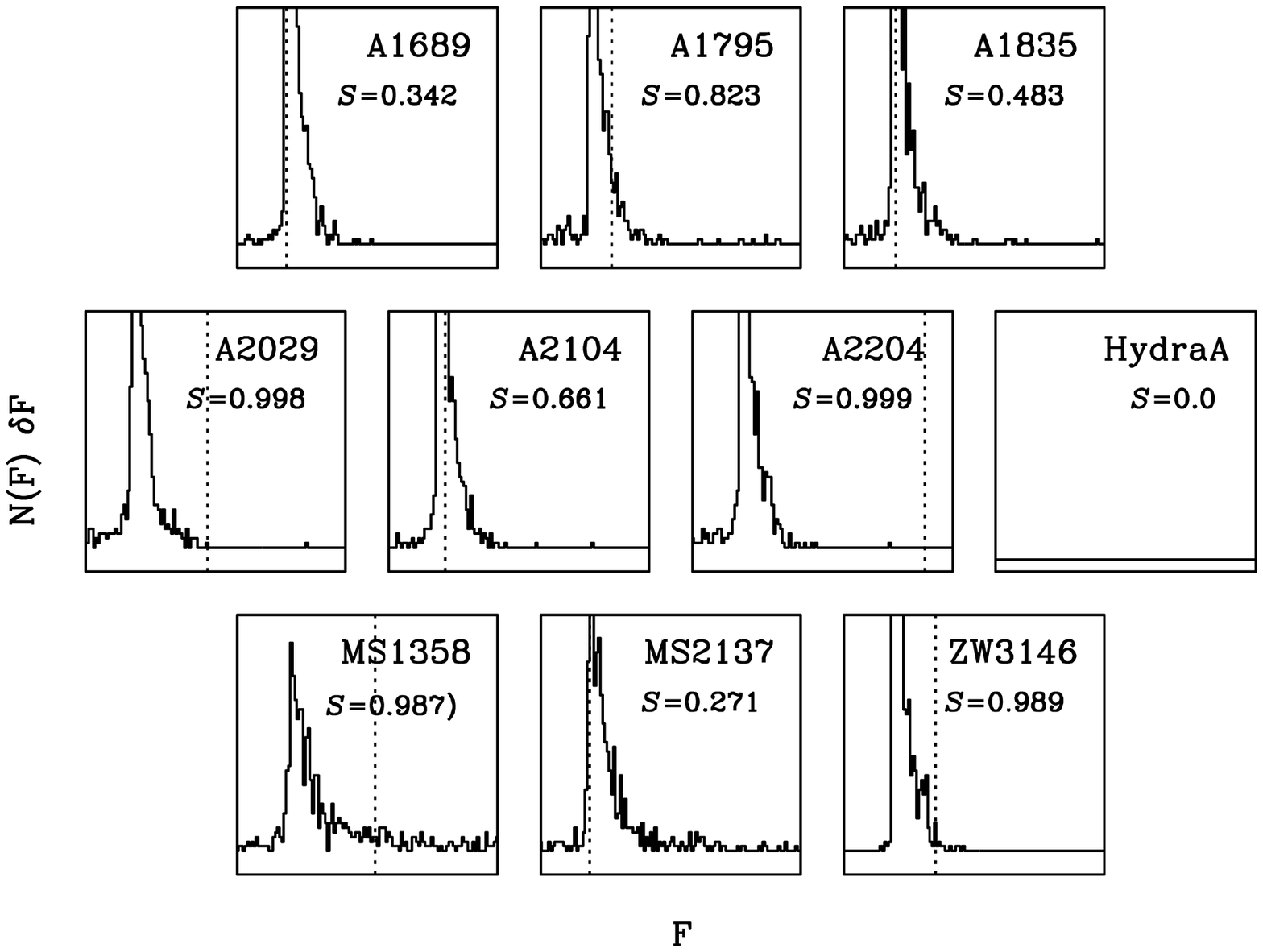}

\figcaption{Empirical $F$ distributions and the MCMC significance $S$ of a
second cospatial core plasma component for each cluster in the sample.  The $F$
value of the original {\it Chandra} data set is denoted with a vertical dashed
line.  (Note that Hydra A data did not admit an $N_c=2$ emission model.)
\label{f15}}


\clearpage
\plotone{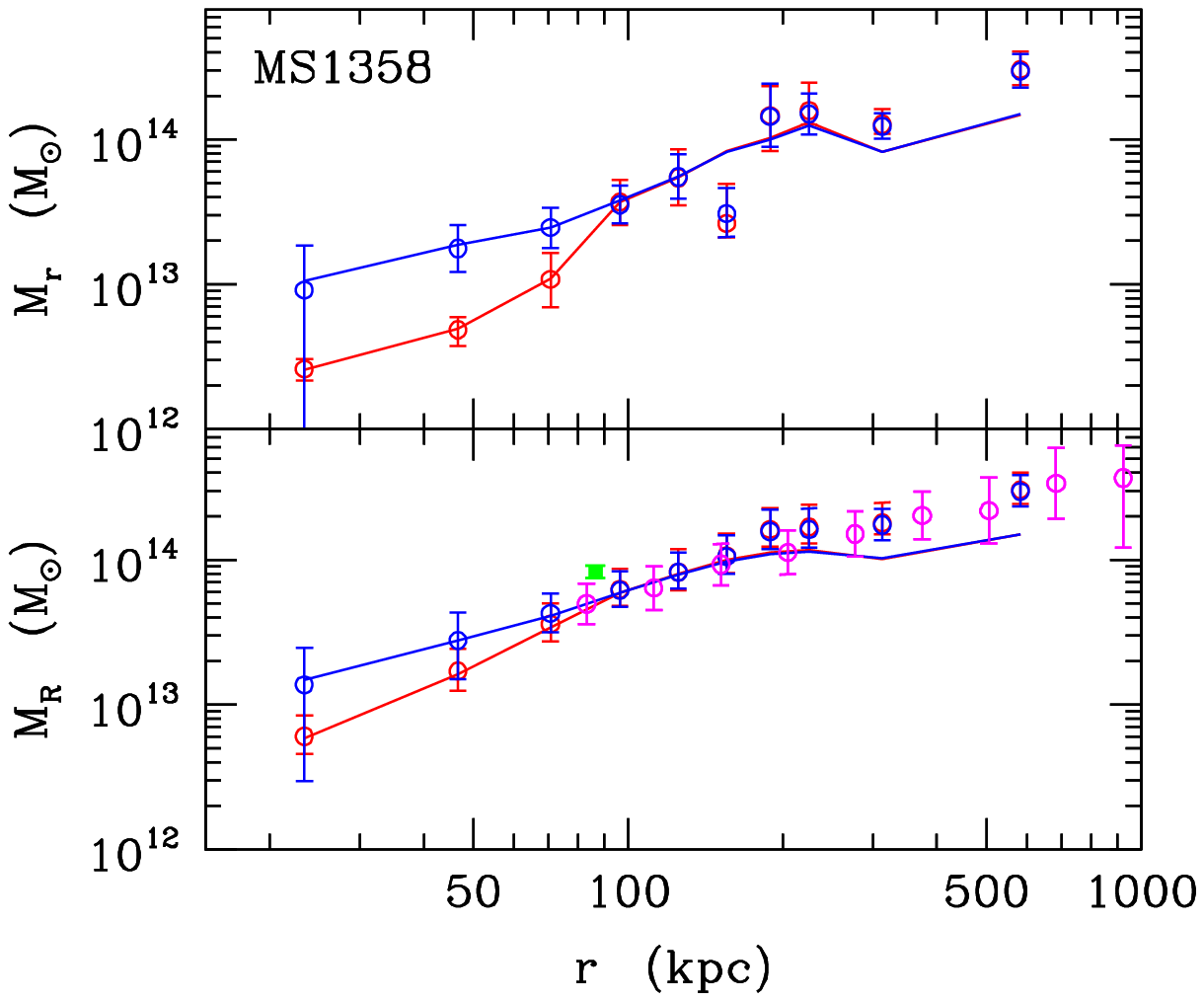}

\figcaption{Mass profiles for models $N_c=0$ (red) and $N_c=2$ (blue) of
MS1358, overlaid for comparison.  Weak (violet) and strong (green)
gravitational lensing measurements are shown for comparison.
\label{f16}}


\clearpage
\plotone{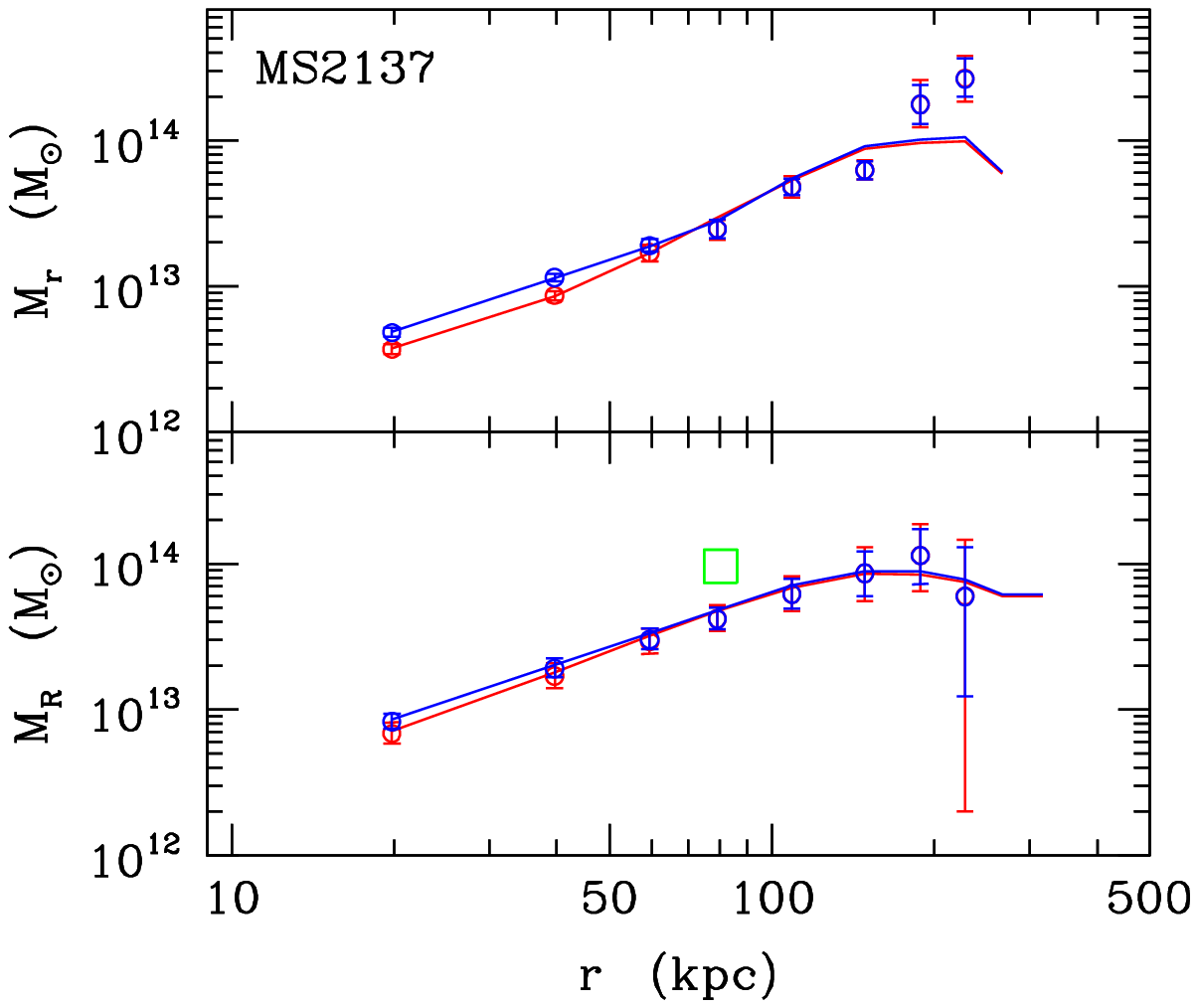}

\figcaption{Same as Figure~\ref{f16}, for MS2137.
\label{f17}}




\clearpage
\begin{center}
\begin{deluxetable}{ll}
\tablewidth{100pt}
\tablecaption{Galaxy Cluster Sample\label{t01}}
\tablehead{
\colhead{cluster} &
\colhead{z}
}
\startdata
A1689  &  0.181   \\
A1795  &  0.0631  \\
A1835  &  0.2523  \\
A2029  &  0.0765  \\
A2104  &  0.1554  \\
A2204  &  0.1523  \\
HydraA &  0.0522  \\
MS1358 &  0.328   \\
MS2137 &  0.313   \\
ZW3146 &  0.2906  \\
\enddata
\end{deluxetable}
\end{center}


\clearpage
\begin{center}
\begin{deluxetable}{llllll}
\tablewidth{400pt}
\tablecaption{Fidelity measures of baryon density ($\chi^2_\rho$) and
temperature ($\chi^2_T$) of each constrained mass profile (see
Equation~\ref{eq12}), and multiphase core plasma significance $S$.
(Note that Hydra A data did not admit an $N_c=2$ emission model.)\label{t02}}
\tablehead{
\colhead{cluster}                 &
\colhead{$\chi^2_\rho$ ($N_c=0$)} &
\colhead{$\chi^2_T$ ($N_c=0$)}    &
\colhead{$\chi^2_\rho$ ($N_c=2$)} &
\colhead{$\chi^2_T$ ($N_c=2$)}    &
\colhead{$S$}
}
\startdata
A1689  & 0.00150  & 0.467 & 0.00825 & 1.14  & 0.342 \\
A1795  & 0.0777   & 7.45  & 0.0156  & 1.29  & 0.823 \\
A1835  & 0.00229  & 1.13  & 1.188   & 1.41  & 0.483 \\
A2029  & 0.00774  & 1.17  & 0.00773 & 1.20  & 0.998 \\
A2104  & 0.00251  & 0.624 & 0.190   & 4.51  & 0.661 \\
A2204  & 0.000994 & 0.759 & 0.00947 & 0.284 & 0.999 \\
HydraA & 0.197    & 2.43  & ---     & ---   & ---   \\
MS1358 & 0.00663  & 0.381 & 0.00454 & 0.356 & 0.987 \\
MS2137 & 0.0101   & 1.08  & 0.0110  & 1.38  & 0.271 \\
ZW3146 & 0.0293   & 1.25  & 0.0150  & 0.913 & 0.989 \\
\enddata
\end{deluxetable}
\end{center}


\clearpage \begin{center}
\begin{deluxetable}{lll}
\tablewidth{410pt}
\tablecaption{Lensing comparisons for the very relaxed cluster
subset.\label{t03}}
\tablehead{
\colhead{cluster} &
\colhead{weak lensing} &
\colhead{strong lensing}
}
\startdata
A1689  & \citet{kcs}      & \citet{wu}          \\
A1835  & \citet{cs}       & \citet{allen}       \\
A2029  & \citet{menard}   & ---                 \\
MS1358 & \citet{hoekstra} & \citet{franx,allen} \\
MS2137 & ---              & \citet{sand}        \\
\enddata
\end{deluxetable}
\end{center}


\end{document}